\documentclass[prd,preprintnumbers,twocolumn,eqsecnum,floatfix,a4paper,nofootinbib,superscriptaddress]{revtex4-1}

\def\mnras{\rm{MNRAS}}

\usepackage{color}
\usepackage{amsmath,amssymb,graphicx,mathtools}
\usepackage{bm,lipsum}
\usepackage{times}
\usepackage{microtype}
\usepackage[utf8]{inputenc}
\usepackage{booktabs}
\usepackage[caption=false]{subfig}
\usepackage{float}
\usepackage[normalem]{ulem}
\usepackage[varg]{txfonts}
\usepackage{bm,graphics,graphicx,epsfig,xcolor,ulem}
\usepackage[colorlinks, pdfborder={0 0 0}]{hyperref}
\usepackage{multirow}
\usepackage{gensymb}
\usepackage{accents}
\definecolor{LinkColor}{rgb}{0.75, 0, 0}
\definecolor{CiteColor}{rgb}{0, 0.5, 0.5}
\definecolor{UrlColor}{rgb}{0, 0, 0.75}
\hypersetup{linkcolor=LinkColor}
\hypersetup{citecolor=CiteColor}
\hypersetup{urlcolor=UrlColor}
\allowdisplaybreaks
\makeatletter
\newcommand*{\rom}[1]{\expandafter\@slowromancap\romannumeral #1@}
\makeatother

\begin{document}
\title{Constraining runaway dilaton models using joint gravitational-wave and electromagnetic observations}


\newcommand{\penncosmos}{\affiliation{Institute for Gravitation and the Cosmos, Department of Physics, Pennsylvania State University, University Park, PA 16802, USA}}
\newcommand{\pennastro}{\affiliation{Department of Astronomy \& Astrophysics, Pennsylvania State University, University Park, PA 16802, USA}}
\newcommand{\cardiff}{\affiliation{School of Physics and Astronomy, Cardiff University, Cardiff, UK, CF24 3AA}
}
\newcommand{\olemiss}{\affiliation{Department of Physics and Astronomy, The University of Mississippi, University, Mississippi 38677, USA}}

\author{Arnab Dhani}
\email{aud371@psu.edu}
\penncosmos 
\author{Anuradha Gupta}
\email{agupta1@olemiss.edu}
\olemiss
\author{B. S. Sathyaprakash}
\email{bss25@psu.edu}
\penncosmos
\pennastro
\cardiff

\begin{abstract}
   With the advent of gravitational-wave astronomy it has now been possible to constrain modified theories of gravity that were invoked to explain the dark energy. In a class of dilaton models, distances to cosmic sources inferred from electromagnetic and gravitational wave observations would differ due to the presence of a friction term. In such theories, the ratio of the Newton's constant to the fine structure constant varies with time. In this paper we explore the degree to which it will be possible to test such models. If collocated sources (e.g. supernovae and binary neutron star mergers), but not necessarily multimessengers, can be identified by electromagnetic telescopes and gravitational-wave detectors one can probe if light and gravitational radiation are subject to the same laws of propagation over cosmological distances. This helps in constraining the variation of Newton's constant relative to fine-structure constant. The next generation of gravitational wave detectors, such as the Cosmic Explorer and Einstein Telescope, in tandem with the Vera Rubin Observatory and gamma ray observatories such as the Fermi Space Observatory will be able to detect or constrain such variations at the level of a few parts in 100. We apply this method to GW170817 with distances inferred by the LIGO and Virgo detectors and the observed Kilonova.
\end{abstract}

\maketitle

\section{Introduction}
Gravitational waves (GWs) and electromagnetic (EM) waves follow the same propagation equations in General Relativity (GR)~\cite{Maggiore:1900zz}. Consequently, the various distance measures in cosmology (e.g., luminosity distance, angular diameter distance, comoving distance, etc.) are identical for both GW and EM. Several alternative theories of gravity with additional scalar degrees of freedom~\cite{Fujii:2003pa,Bertolami:2007gv,Bertolami:2008ab, Bertolami:2008im,Sotiriou:2008it,DeFelice:2010aj,Harko:2012hm,Das:2008iq,Bisabr:2012cu,Moffat:2010ek,Shiralilou:2021mfl} modify the propagation of either or both by altering the friction term in the wave equations due to the evolution of the scalar field. We will, however, restrict to a class of scalar-tensor theories in which the dispersion relation remains unchanged. 
Hence, in these scalar-tensor theories distance to an astronomical source inferred from gravitational-wave observation will be different from that inferred using electromagnetic radiation.

The presence of a scalar field is also motivated by the low energy effective field theories of Loop Quantum Gravity~\cite{Rovelli:1993bm, Domagala:2010bm} and String Theory~\cite{Green:1987mn,Uzan:2010pm,Damour:1994ya,Damour:1994zq,Gasperini:2001pc,Minazzoli:2013ara}. Furthermore, dark energy~\cite{Ratra:1987rm,Caldwell:1997ii,Peebles:2002gy}, inflation~\cite{Guth:1980zm,Linde:1981mu,Albrecht:1982wi,Linde:2007fr}, and variations of the fundamental constants are often modeled using a scalar field~\cite{Bekenstein:1982eu,Sandvik:2001rv,Dvali:2001dd,Olive:2007aj,Damour:2012rc}. In fact, it has been claimed that the requirement of gauge and diffeomorphism invariances would invariably lead to scalar-tensor theories with minimal/non-minimal coupling to the matter sector~\cite{ArmendarizPicon:2002qb}. The coupling of the scalar field to the gravitational sector in such theories has been tightly constrained in the weak-field limit using solar system tests~\cite{Adelberger:2003zx,Adelberger:2006dh,Adelberger:2009zz,Kapner:2006si,Will:2005va}. If the scalar field couples non-minimally to the matter sector, the Einstein equivalence principle is broken. The equivalence principle, likewise, has been tested to a very high accuracy within the solar system~\cite{Rosenband1808,Will:2005va,Adelberger:2009zz,Williams:2012nc}. A variety of decoupling~\cite{Tseytlin:1991xk,Damour:1995pd,Damour:2002mi,Jarv:2008eb,Damour:1992kf,Damour:1994zq,Minazzoli:2013ara} or screening~\cite{Khoury:2010xi,Khoury:2003aq,Khoury:2003rn,Hees:2011mu,Hinterbichler:2010es,Hinterbichler:2011ca} mechanisms have, therefore, been proposed to keep these theories viable for cosmological evolution. 

The propagation of waves on a modified background allows one to test for the presence of a scalar field on cosmological scales. High redshift quasar absorption spectra~\cite{Webb:2000mn,King:2012id,Webb:2010hc}, galaxy clustering data~\cite{Holanda:2015oda}, and 21cm neutral hydrogen intensity mapping~\cite{Khatri:2007yv} have been used to place limits on the spatio-temporal evolution of the fine structure constant which can be modeled using a scalar field. Type Ia supernova (SNeIa) data is used to fit the EM luminosity distance-redshift relation and constrain models of dynamical dark energy~\cite{Riess:2016jrr}. Other studies use the EM luminosity distance estimates from SNeIa in parallel with the EM angular diameter distance measurements from X-ray and Sunyaev-Zel'dovich observations of galaxy clusters to directly constrain the violation of the distance-duality relation in the EM sector~\cite{2011RAA....11.1199C,Hees:2014lfa}. 

Gravitational wave astronomy has opened a new means of revealing the presence of a scalar degree of freedom. Coincident measurements of the luminosity distance from GW observations and the redshift from follow-up EM observations of ``bright" sirens, such as the first observation of gravitational waves from a binary neutron star merger, GW170817~\cite{LIGOScientific:2017vwq,LIGOScientific:2017ync}, have been used to put limits on the modified friction term in $f(R)$ and scalar-tensor theories of gravity~\cite{Fanizza:2020hat,Finke:2021aom}. The luminosity distance-redshift relation has also been constrained for ``dark" sirens (GW observations without an EM counterpart) by cross-correlations with galaxy catalogs~\cite{Finke:2021aom,Mukherjee:2020mha}. In these methods, the modified friction term is constrained together with the standard cosmological parameters, however, the two sets of parameters are strongly correlated with each other. \textcite{Mukherjee:2020mha} propose the use of Baryon acoustic oscillation (BAO) data together with luminosity-distance measurements from GW observations and redshifts from galaxy catalog cross-correlations to directly constrain the ratio between the GW luminosity-distance and EM luminosity distance in terms of the modified friction parameter.

In this study, we propose the direct use of the EM luminosity distance from SNeIa/kilonova concurrently with the GW luminosity distance from ``bright" sirens and the redshift obtained from photometric/spectroscopic studies of the identified galaxy or galaxy cluster to directly constrain the ratio of the two luminosity distances for a class of scalar-tensor theories with a non-minimal multiplicative coupling between the scalar field and the matter sector. The crucial distinction with the method described in~\textcite{Mukherjee:2020mha} is their use of the BAO data to convert the angular diameter distance to EM luminosity distance via the distance-duality relation, which is broken for us due to the non-minimal coupling of the scalar field to the matter sector. In other words, their procedure is valid for alternative theories of gravity in which gravity is minimally coupled to the matter sector whereas our method applies to more general theories. Furthermore, they infer the redshifts to GW sources using galaxy correlation and as a result also measure some cosmological parameters. We restrict ourselves to ``bright" sirens and, therefore, have a direct measurement of the redshift. In this way, our parameter constraints do not suffer from degeneracies with the other cosmological parameters.

The class of scalar-tensor theories considered in this study arise as low energy action of string theories and satisfy the solar system tests for both the modifications to the gravitational sector and the breakage of the equivalence principle. This class of theories, known as the runaway dilaton models~\cite{Gasperini:2001pc,Damour:2002mi,Minazzoli:2014xua,Hees:2014lfa}, has a Brans-Dicke type gravitational interaction and a universal multiplicative coupling between the scalar field and the matter sector which breaks the equivalence principle. The unequal coupling of the scalar field to the metric and the matter sector leads to (distinct) modified propagation equations for gravitational waves and electromagnetic waves. 

We parameterize the ratio of the electromagnetic and gravitational-wave luminosity distance using a parameter $\eta_0$. We find that the planned upgrades to the second-generation of advanced gravitational-wave detector networks (e.g. the A+ upgrade  \cite{Reitze:2019iox,KAGRA:2013rdx} of Advanced LIGO and similar upgrades to Advanced Virgo \cite{VIRGO:2014yos},  KAGRA \cite{Aso:2013eba,Somiya:2011np} and LIGO-India \cite{Unnikrishnan:2013qwa,Saleem:2021iwi}) constrains $\eta_0$ to $|\eta_0|<0.2,$ while the proposed improvement of the network to Voyager sensitivity \cite{LIGO:2020xsf} refines the constraint to $|\eta_0|<0.05$. The proposed third-generation of ground-based gravitational-wave detector network (Cosmic Explorer \cite{Evans:2021gyd,Reitze:2019iox} and Einstein Telescope \cite{Punturo:2010zz,Punturo:2010zza,Hild:2010id}) will place the best limits on $\eta_0$ at $|\eta_0|<0.01$.

In Sec.~\ref{sec:background}, we briefly describe runaway dilaton models and their EM and GW propagation equations. We also describe how the ratio of the luminosity distance of each sector can be related to the variation of the fundamental constants. In Sec.~\ref{sec:method}, we discuss the gravitational-wave detectors considered in this study, the simulations we performed, and the electromagnetic data that we used. We outline our main results and forecasts in Sec.~\ref{sec:results} and the constraints that can be placed using GW170817 in Sec.~\ref{sec:gw170817_constraints}. Sec.~\ref{sec:conclusion} concludes the paper.

\section{Background}
\label{sec:background}
In this section, we briefly review the equations of motion for runaway dilaton models, derive the propagation equations for electromagnetic and gravitational waves on a homogeneous and isotropic background, parameterize the ratio of the luminosity distances as a function of redshift, and relate it to the redshift variation of the fundamental constants.

\subsection{Runaway dilaton models}
The action for runaway dilaton models~\cite{Gasperini:2001pc,Damour:2002mi,Minazzoli:2014xua,Hees:2014lfa}, a class of scalar-tensor theories with a generic multiplicative coupling $h(\phi)$ between a scalar field $\phi$ and the matter Lagrangian $\mathcal{L}_m[g_{\mu\nu,\Psi}]$, is given by
\begin{multline}
\label{eq:action}
    S = \int d^4x \sqrt{-g} \Bigg[ \frac{1}{2\kappa} \left( \phi R - \frac{\omega(\phi)}{\phi} \nabla_{\mu}\phi\nabla^{\mu}\phi - V(\phi) \right) \\
    + h(\phi)\mathcal{L}_m[g_{\mu\nu,\Psi}] \Bigg]\,,
\end{multline}
where $\kappa=8\pi G$ with $G$ being the gravitational coupling constant, $R$ is the Ricci scalar, and $\Psi$ consists of all the Standard Model fields.

The gravitational equations of motion, given by the variation of the action with respect to the metric, takes the form,
\begin{multline}
\label{eq:gr}
    R_{\mu\nu}-\frac{1}{2}g_{\mu\nu}R = \kappa \frac{h(\phi)}{\phi}T_{\mu\nu} + \frac{1}{\phi}(\nabla_{\mu}\nabla_{\nu}-g_{\mu\nu}\Box)\phi \\ 
    + \frac{\omega(\phi)}{\phi^2}(\nabla_{\mu}\phi\nabla_{\nu}\phi-\frac{1}{2}g_{\mu\nu}\nabla_{\alpha}\phi\nabla^{\alpha}\phi) - g_{\mu\nu}\frac{V(\phi)}{2\phi}.
\end{multline}
Similarly, one can obtain the equation of motion for the scalar field by varying the action with respect to it. Upon replacing the Ricci scalar in the resulting equation with the trace of the gravitational equations of motion Eq.~(\ref{eq:gr}), one finds that the Klein-Gordon equation for the scalar field is given by,
\begin{multline}
\label{eq:scalar}
    \frac{2\omega(\phi)+3}{\phi}\Box\phi = \kappa\left(\frac{h(\phi)}{\phi}T^{\alpha}_{\alpha}-2h'(\phi)\mathcal{L}_m\right) \\
    -\frac{\omega'(\phi)}{\phi}\nabla_{\alpha}\phi\nabla^{\alpha}\phi + V'(\phi) - 2\frac{V(\phi)}{\phi},
\end{multline}
where a prime denotes the derivative with respect to $\phi$. The stress-energy tensor $T_{\mu\nu}$ in the above equations can be defined by the variation of the matter Lagrangian with the metric $g_{\mu\nu}$,
\begin{equation}
    T_{\mu\nu} \equiv \frac{-2}{\sqrt{-g}} \frac{\delta}{\delta g^{\mu\nu}} (\sqrt{-g}\mathcal{L}_m).
\end{equation}

We will consider the background to be a homogeneous, isotropic, and spatially flat Universe described by the \textit{Friedmann-Lema\^itre-Robertson-Walker} (FLRW) metric,
\begin{equation}
\label{eq:frlw}
    ds^2 = -dt^2 + a(t)^2 \delta_{ij} dx^i dx^j,
\end{equation} 
where the size of the homogeneous, isotropic, and spatially flat 3-surface is given by the scale factor $a(t)$. The background spacetime is considered to be sourced by a perfect fluid with its stress-energy tensor given by
\begin{equation}
\label{eq:fluid}
    T^{\mu\nu} = (\rho+p)u^{\mu}u^{\nu} + pg^{\mu\nu},
\end{equation}
where $\rho$ is the total energy density and $p$ is the pressure of the fluid in its rest-frame, and $u^{\mu}$ is the 4-velocity of the fluid with respect to an observer. 

The Friedmann equations that describe the evolution of the background spacetime are obtained by substituting Eq.~(\ref{eq:frlw}) and Eq.~(\ref{eq:fluid}) in to the gravitational field equations (\ref{eq:gr}): 
\begin{equation}
\label{eq:fried1}
    H^2 = \kappa\frac{h(\phi)}{3\phi} \rho + \frac{V(\phi)}{6\phi} + \frac{\omega(\phi)}{6}\left(\frac{\dot{\phi}}{\phi}\right)^2 - H\frac{\dot{\phi}}{\phi},
\end{equation}
\begin{equation}
\label{eq:fried2}
    2\dot{H}+3H^2 = - \kappa\frac{h(\phi)}{\phi}p + \frac{V(\phi)}{2\phi} -2H\frac{\dot{\phi}}{\phi} - \frac{\omega(\phi)}{2}\left(\frac{\dot{\phi}}{\phi}\right)^2 - \frac{\ddot{\phi}}{\phi},
\end{equation}
where $H(t)$ is the Hubble parameter defined as $H(t)\equiv\dot{a}(t)/a(t)$ and dots denote derivatives with respect to the time coordinate $t$. In Eqs.~(\ref{eq:fried1}) and~(\ref{eq:fried2}), if the scalar field $\phi$ is a constant, only the first two terms on the right-hand side are non-zero and we recover the standard Friedmann equations for $\Lambda$CDM cosmology up to the normalization of the field $\phi$.

\subsection{Propagation of gravitational waves}
In this and the following subsection, we will derive the equations describing the propagation of gravitational and electromagnetic waves, respectively, on the background spacetime.

Gravitational waves are propagating tensor perturbations of the background spacetime. To get the equations of motion for tensor perturbations, we perturb our FLRW metric as
\begin{equation}
    ds^2 = -dt^2 + a(t)^2 (\delta_{ij}+h_{ij}) dx^i dx^j,
\end{equation}
where $h_{ij}$ is a small perturbation of the background geometry and is transverse ($\partial_i h^{ij}=0$) and trace-less ($h_i^i=0$) in the chosen coordinate system. Note that this is not a generic perturbation of the background. A generic perturbation can be decomposed in to scalar, vector, and tensor components that do not mix under diffeomorphisms. Furthermore, at the leading order, the equations of motion for these components are decoupled. Here, since we are only interested in GWs, it is sufficient to perturb the background with the tensor component which is transverse and trace-less.

The equations of motion for gravitational-wave propagation are then given by
\begin{equation}
\label{eq:gw}
    \ddot{h}_{ij} + \left( 3H+\frac{\dot{\phi}}{\phi} \right)\dot{h}_{ij} - \frac{\nabla^2h_{ij}}{a(t)^2} =0,
\end{equation}
where $\dot{\phi}/\phi$ is the modified friction term that would change the observed GW amplitude and as a result the luminosity distance with respect to GR. We note that the luminosity distance is additionally modified since the Friedmann equations get altered due to the presence of the scalar field. In other words, the evolution of the scale factor $a(t)$ is different from that in GR. Note, however, that the dispersion relation is unchanged with respect to GR and, hence, GWs travel at the \textit{speed of light}.

Throughout this study, we are interested in solutions of the wave equations under geometric optics approximation. This is because the length scales of the signals of interest to us (stellar-mass compact binary mergers and SNeIa) are much smaller than the Hubble scale. In this limit, the metric perturbations can be written as
\begin{equation}
\label{eq:gw_wkb}
    h_{ij} = \mathcal{R} \{(b_{ij}+\epsilon c_{ij}+\mathcal{O}(\epsilon^2)) \,e^{i \theta/\epsilon}\},
\end{equation}
where $\theta$ is the phase of the plane wave and $\epsilon$ is an order-counting parameter, which can be set to 1 at the end of the calculation. Substituting Eq.~(\ref{eq:gw_wkb}) into Eq.~(\ref{eq:gw}) and collecting terms of the same order, we get
\begin{equation}
\begin{aligned}
    k_{\mu}k^{\mu} &= 0, \\ 
    k^{\nu}\nabla_{\nu}k^{\mu} &= 0,
\end{aligned}
\end{equation}
at the $\mathcal{O}(\epsilon^2)$ and 
\begin{equation}
    \nabla_{\mu}(b^2 k^{\mu}) = -b^2 k^{\mu}\nabla_{\mu} \ln \phi,
\end{equation}
at the $\mathcal{O}(\epsilon)$, where $k_{\mu}=\nabla_{\mu}\theta$, the wave vector, is null and follows null geodesics and $b=||b_{ij}||$ is the Euclidean norm of the leading-order amplitude. The latter equation is the one which is modified with respect to GR and represents the non-conservation of the graviton number as it propagates on the background spacetime.

The luminosity distance can then be calculated following \textcite{Minazzoli:2014xua}\footnote{The derivation of the luminosity distance is carried out for EM waves but the procedure is the same for GWs.} and is given by
\begin{equation}
    d_L^{\rm GW} = (1+z) \sqrt{\frac{\phi_0}{\phi}} \int_0^z \frac{dz}{H(z)},
\end{equation}
where where $\phi_0$ is the value of the field in the present epoch and $H(z)$ is the modified Hubble relation [Eq.~(\ref{eq:fried1})].

\subsection{Propagation of electromagnetic waves}
The field equations that govern the propagation of electromagnetic waves can be obtained by the variation of the action Eq.~(\ref{eq:action}) with respect to the EM 4-potential $A^{\mu}$ and are given by
\begin{equation}
    \nabla_{\nu}(h(\phi)F^{\mu\nu}) = 0,
\end{equation}
where $F_{\mu\nu}=\nabla_{\mu}A_{\nu}-\nabla_{\nu}A_{\mu}$ is the electromagnetic field tensor. 

Solving the above equation in the geometric optics limit, as for metric perturbations, yields similar equations for photons, namely, they travel on null geodesics. The equation for photon number `non-conservation' is given by:
\begin{equation}
    \nabla_{\mu}(\overline b^2 k^{\mu}) = -\overline b^2 k^{\mu}\nabla_{\mu} \ln h(\phi),
\end{equation}
where the electromagnetic potential in the geometric optics limit is given by
\begin{equation}
    A^{\mu}=\mathcal{R}\{ (b^{\mu}+\epsilon c^{\mu}+\mathcal{O}(\epsilon^2))\,e^{i \theta/\epsilon} \},
\end{equation}
with $\overline b=||b^{\mu}||$ being the norm taken with respect to the background FLRW metric.

The luminosity distance is then given by~\cite{Minazzoli:2014xua}
\begin{equation}
    d_L^{\rm EM} = (1+z) \sqrt{\frac{h(\phi_0)}{h(\phi)}} \int_0^z \frac{dz}{H(z)}\,,
\end{equation}
where $\phi_0$ is the value of the field at the present epoch and $H(z)$ is given by the modified Friedman equations. Of note here is that the two luminosity distances differ only if $h(\phi)\neq \phi$ which is the premise we are working under.

\subsection{Parameterizing the modified luminosity distances}
Now that we have obtained the luminosity distance-redshift relation for both electromagnetic and gravitational waves, we can parameterize the scalar field dependence of their ratio. We choose to do this parameterization in terms of the violation of the distance-duality relation for both the sectors. This choice helps connect our results to those of the numerous experiments in the electromagnetic sector that parameterize this deviation~\cite{Hees:2014lfa}. The distance-duality relation connects the luminosity distance to the angular diameter distance. The latter is defined by:
\begin{equation}
    d_A(z) = \frac{1}{1+z} \int_0^z \frac{dz}{H(z)}.
\end{equation}
This is a geometric quantity that can be derived by integrating the geodesic equation. For the class of theories considered in this study, both gravitational and electromagnetic waves travel on null geodesics and, therefore, their angular diameter distances are unchanged from that of GR, apart from a modification to the Friedmann equations. We can, then, write the parameterization as
\begin{equation}
    \eta(z) = \frac{d_L(z)}{d_A(z)(1+z)^2}.
\end{equation}
In GR, the distance-duality relation implies $\eta(z)=1$. 

We consider $\eta(z)$ to have the functional form
\begin{equation}
\begin{aligned}
    \eta^{\rm GW}(z) = \sqrt{\frac{\phi_0}{\phi}} = 1 + \eta_{1}\frac{z}{1+z}, \\
    \eta^{\rm{} EM}(z) = \sqrt{\frac{h(\phi_0)}{h(\phi)}} = 1 + \eta_{2}\frac{z}{1+z},
\end{aligned}
\end{equation}
parameterizing the deviation from the gravitational (electromagnetic) distance-duality relation by $\eta_1$ ($\eta_2$). Additionally, we parameterize the ratio of the luminosity distances as
\begin{equation}
    \frac{d_L^{\rm GW}}{d_L^{\rm EM}} = \frac{\eta^{\rm GW}(z)}{\eta^{\rm EM}(z)} = \sqrt{\frac{\phi_0/h(\phi_0)}{\phi/h(\phi)}} = 1 + \eta_0\frac{z}{1+z}
\end{equation}
from which one can deduce that $\eta_0 \approx \eta_1-\eta_2$. The above form of the parameterization was introduced in~\textcite{Holanda:2012at} with the advantage being that it avoids divergence at large redshifts which the linear expansions suffer from.
Given a simultaneous measurement of the GW and EM luminosity distances, either from the same source or the same galaxy or the same galaxy cluster (see Sec.~\ref{subsubsec:sne_rates} and \ref{sec:conclusion} for a discussion), one can place constraints on the parameter $\eta_0$. 

At this point, we note that studies in~\cite{Mukherjee:2020mha,Fanizza:2020hat,Finke:2021aom} have constrained the ratio of the two luminosity distances, albeit in the context of modifying the frictional term in the gravitational sector alone, through the parameterization,
\begin{equation}
    \frac{d_L^{\rm GW}(z)}{d_L^{\rm EM}(z)} = \Xi_0 + \frac{1-\Xi_0}{(1+z)^n}.
\end{equation}
where $\Xi_0=1$ in GR and $n$ gives the rate at which the ratio saturates to its asymptotic value $\Xi_0$. Our parameter $\eta_0$ is related to $(\Xi_0,n)$ as
\begin{equation}
    \eta_0 = 1 - \Xi_0 \quad {\rm for} \quad n=1,
\end{equation}
i.e., our parameterization is a subclass of the $(\Xi_0,n)$ parameterization for a fixed saturation rate.


The errors on $\eta_0$ can be calculated from the errors on the GW and EM luminosity distances, assuming the redshift to the source is known, using the standard error propagation formula for independent variables ($d_L^{\rm GW}$ and $d_L^{\rm EM}$ in this case) as,
\begin{equation}
    \sigma_{\eta_0}^2 = \left(\frac{\partial \eta_0}{\partial d_L^{\rm GW}}\right)^2 \sigma_{d_L^{\rm GW}}^2 + \left(\frac{\partial \eta_0}{\partial d_L^{\rm EM}}\right)^2 \sigma_{d_L^{\rm EM}}^2,
\end{equation}
where $\sigma_X$ denotes $1$-$\sigma$ error in the quantity $X$. Simplifying the above equation by evaluating the derivative expressions leads to
\begin{equation}
\label{eq:eta_err}
    \sigma_{\eta_0} = \frac{1+z}{z} \frac{d_L^{\rm GW}}{d_L^{\rm EM}} \sqrt{\left(\frac{\sigma_{d_L^{\rm GW}}}{d_L^{\rm GW}}\right)^2 + \left(\frac{\sigma_{d_L^{\rm EM}}}{d_L^{\rm EM}}\right)^2}.
\end{equation}

\subsection{Redshift variation of fundamental constants}
The non-minimal coupling of the scalar field to the matter and gravitational sectors leads to the dependence of the fundamental constants on the scalar field and they, therefore, evolve with the evolution of the scalar field~\cite{Yunes:2009bv,Yunes:2016jcc,Vijaykumar:2020nzc}. From the action given by Eq.~(\ref{eq:action}), it can be read out that the fine structure constant $\alpha$ and the gravitational constant $G$ depend on the scalar field via
\begin{equation}
\begin{aligned}
    \alpha \sim h^{-1}(\phi), \\
    G \sim \phi^{-1},
\end{aligned}
\end{equation}
and, hence, their redshift variation can be written as
\begin{equation}
\begin{aligned}
    \frac{\Delta\alpha(z)}{\alpha_0} \equiv \frac{\alpha(z)-\alpha_0}{\alpha_0} = \frac{h(\phi_0)}{h(\phi)} - 1 = \eta^{\rm EM}(z)^2-1 \\
    \frac{\Delta G(z)}{G_0} = \frac{G(z)-G_0}{G_0} = \frac{\phi_0}{\phi}-1 = \eta^{\rm GW}(z)^2-1\,,
\end{aligned}
\end{equation}
where $\alpha_0$ and $G_0$ are the values of $\alpha$ and $G$ at the current epoch, respectively. 

Given the experimental constraints on the ratio of the two luminosity distances, one can constrain the temporal variation of $\frac{G}{\alpha}(z)$ in the current epoch as 
\begin{equation}
\label{eq:temp_var}
    \beta \equiv \left.\frac{\tfrac{d}{dt}(G/\alpha)}{G/\alpha}\right\vert_0 = -2 H_0 \left.\frac{d\eta}{dz}\right\vert_0 = -2 H_0 \eta_0\,,
\end{equation} 
where $\eta(z)=\eta^{\rm GW}(z)/\eta^{\rm EM}(z)$ and $H_0$ is the present value of the Hubble parameter. If one uses constraints from other electromagnetic probes~\cite{Hees:2014lfa}, the temporal variations of both $\alpha$ and $G$ can be separately constrained.

We point out here that $G_0$ is not the effective gravitational constant $G_{\rm eff}$ that enters the Poisson equation at the Newtonian order and should not be interpreted as the strength of the gravitational force between two test masses separated by a unit distance. The two are related by
\begin{equation}
    G_{\rm eff} = G_0 \left(1 + \frac{1-2\phi_0 \frac{h'(\phi_0)}{h(\phi_0)}}{2\omega(\phi_0)+3}\right) \frac{h(\phi_0)}{\phi_0}.
\end{equation}
In the absence of the scalar field, $G_0$ and $G_{\rm eff}$ coincide, as expected.



\section{Method}
\label{sec:method}
In this section, we describe the different gravitational wave detector networks considered in this study, outline our procedure for simulating gravitational-wave sources, calculate the rate of spatially coincident EM and GW signals, and estimate the distribution of luminosity distance errors for the coincidentally observed population of sources.

\subsection{Gravitational wave detector networks}
\begin{figure}[h]
    \centering
    \includegraphics[width=\columnwidth]{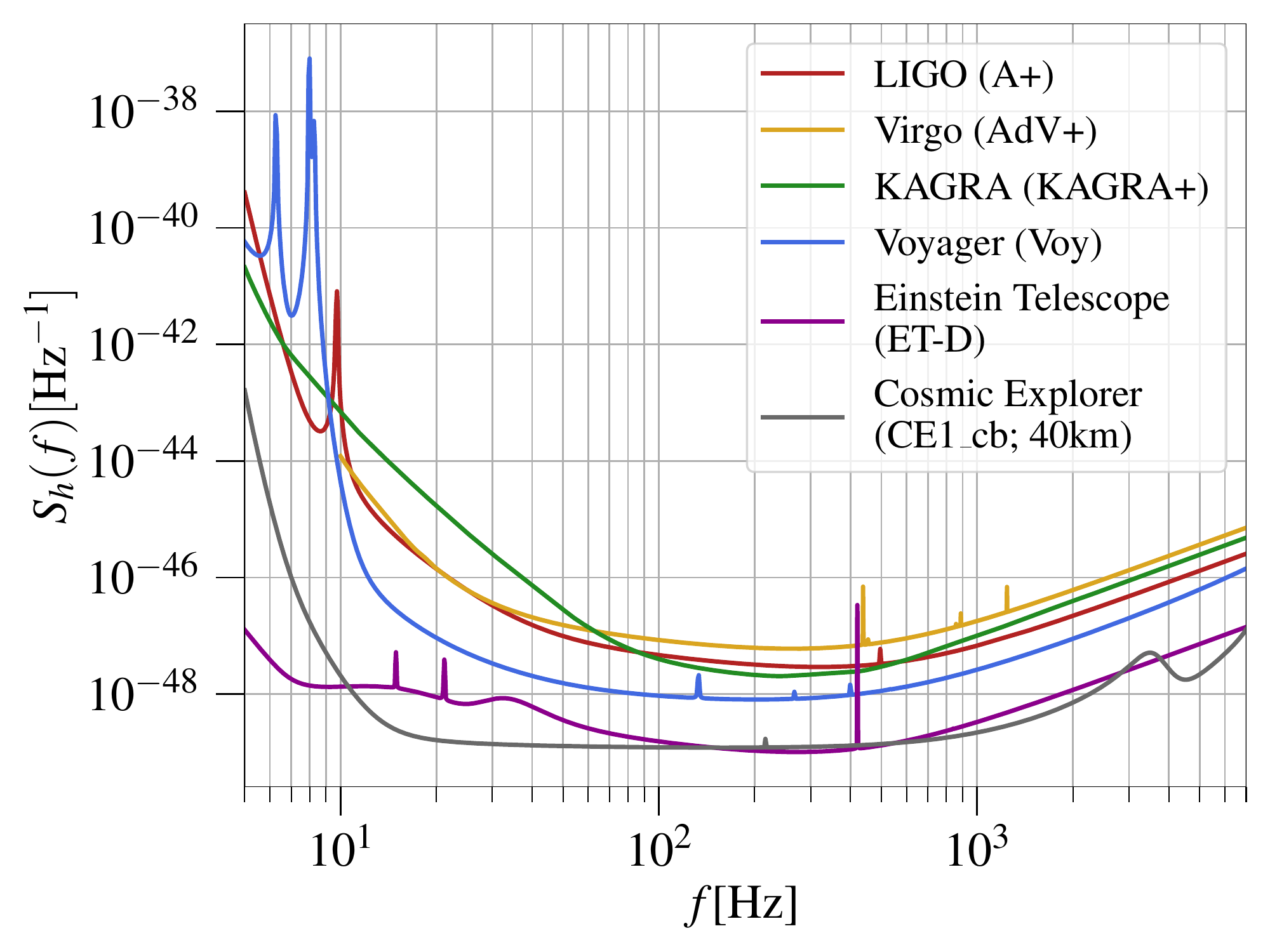}
    \caption{The noise power spectral density (PSD) estimates used for the individual detectors considered in this study. We use a low frequency cutoff of 5Hz for all but the advanced Virgo detector for which the PSD starts at 10Hz.}
    \label{fig:psd}
\end{figure}

We consider three GW detector networks across three technology generations. The \textit{2G+} network consists of the five second generation GW detectors with three LIGO detectors~\cite{TheLIGOScientific:2014jea} (LIGO-Hanford, LIGO-Livingston, LIGO-India) operating at \textit{A+} sensitivity, the Virgo~\cite{TheVirgo:2014hva} and the KAGRA~\cite{Akutsu:2018axf} detectors at \textit{AdV+} and \textit{KAGRA+} sensitivities, respectively.
The \textit{Voy+} network includes the same five second generation detectors but with the LIGO detectors upgraded to a proposed `Voyager'~\cite{Adhikari:2020gft} technology. The final network, \textit{ECC}, consists of three proposed third generation detectors, specifically, two Cosmic Explorer~\cite{Reitze:2019iox} detectors and an Einstein Telescope~\cite{Punturo:2010zza}. We show the noise power spectral densities (PSDs) for the individual detectors in Fig.~\ref{fig:psd}. The locations of these detectors and the technologies used in a network are given in Table~\ref{tab:networks}.

\begin{table}[h]
\centering
    \begin{tabular}{p{0.25\columnwidth}|p{0.70\columnwidth}}
    \hline \hline 
    Network label & Detector location (technology) \\
    \hline
    \hline
    \textit{2G+} & Hanford WA (A+), Livingston LA (A+), Cascina Italy (AdV+), Kamioka Japan (KAGRA+), Hingoli India (A+) \\
    \hline
    \textit{Voy+} & Hanford WA (Voyager), Livingston LA (Voyager), Cascina Italy (AdV+), Kamioka Japan (KAGRA+), Hingoli India (Voyager) \\
    \hline
    \textit{ECC} & Cascina Italy (ET-D), fiducial US site (CE1\_cb), fiducial Australian site (CE1\_cb) \\
    \hline \hline
    \end{tabular}
    \caption{An overview of the three networks used in the study. The location determines the detector antenna patterns, while the technology indicates the used power spectral density. The Voyager and Cosmic Explorer power spectral densities are chosen to be low-frequency optimized and in the case of the latter for a detector arm length of $40\,\text{km}$.}
    \label{tab:networks}
\end{table}

\subsection{Rates}
\label{subsec:rates}
\subsubsection{Binary neutron star merger rates}
We simulate a population of binary neutron star (BNS) merger events up to a redshift of $z=1.5$ assuming a uniform mass distribution between 1$M_{\odot}$ and 2.5$M_{\odot}$ for the individual NSs~\cite{Abbott:2020gyp}. The other parameters, cosine of the inclination angle $\cos \iota$, location of the source on the plane of the sky $\Omega$ (cosine of the declination angle $\cos\delta$ and right ascension $\alpha$), polarization angle $\psi$, and the phase of the signal at coalescence $\phi_0$, of the fiducial BNS population are drawn from a uniform distribution across their domains. We assume 10 years of observing time for each network with an 80\% duty cycle for each detector~\cite{Belgacem:2019tbw}. The redshift distribution for our BNS population is given by the following probability distribution,
\begin{equation}
    p(z) = \frac{R_z(z)}{\int_0^{10}R_z(z)dz},
\end{equation}
where an upper limit of $z=10$ is justified since the contribution to the integral from redshifts larger than 10 is negligible. 
$R_z(z)$, the merger rate density in the observer frame, can be expressed as
\begin{equation}
    R_z(z) = \frac{R_m(z)}{1+z} \frac{dV(z)}{dz}.
\end{equation}
Here $R_m(z)$ is the merger rate per comoving volume in the source frame and $dV/dz$ is the comoving volume element. The former is given by
\begin{equation}
    R_m(z) = \int_{t_{\rm min}}^{t_{\rm max}} R_f[t(z)+t_d]P(t_d)dt_d,
\end{equation}
where $R_f(t)$ is the binary star formation rate (SFR) which we assume follows the Vangioni cosmic SFR~\cite{Vangioni:2014axa}. The delay time (the time it takes for a binary to coalesce after formation) distribution is taken to be $P(t_d) \propto 1/t_d$ with $t_{\rm min} = 20 \,\rm Myr$ and $t_{\rm max}$ set to the Hubble time $1/H_0$. The value of $R_m$ at $z=0$ is estimated from the population properties of the third LIGO--Virgo Gravitational-Wave Transient Catalog, \textit{GWTC-3}~\cite{LIGOScientific:2021psn} to be between
\begin{equation}
\label{eq:local_merger_rate}
    R_m(z=0) = \mbox{13--1900} \; \rm Gpc^{-3}yr^{-1}.
\end{equation}
We present results for both the optimsitic and pessimistic local merger rates.

\subsubsection{Electromagnetic counterpart}
\label{sec:rate_em}
The \textit{Voy+} and the \textit{ECC} network of GW detectors have a reach beyond the horizon distance of the current and future EM telescopes for kilonova which can be observed up to a redshift of about $z=0.5$ (see, e.g. Table 2.2 in Ref.\,\cite{Kalogera:2021bya}). Hence, BNS events beyond a redshift of $z=0.5$ is electromagnetically observable only through short gamma ray burst events. Therefore, in this study, we assume that 10\% of the BNS events up to a redshift of 0.5 will have a dedicated EM follow-up search to detect their kilonova emissions (we assume this to be in addition to possible GRB detection, which do not need a dedicated search owing to the near all-sky sensitivity of GRB detectors) and for BNS observations beyond a redshift of 0.5, we assume a coincident electromagnetic detection to consist of only GRBs. 

\begin{figure*}[ht]
    \centering
    \includegraphics[width=2\columnwidth]{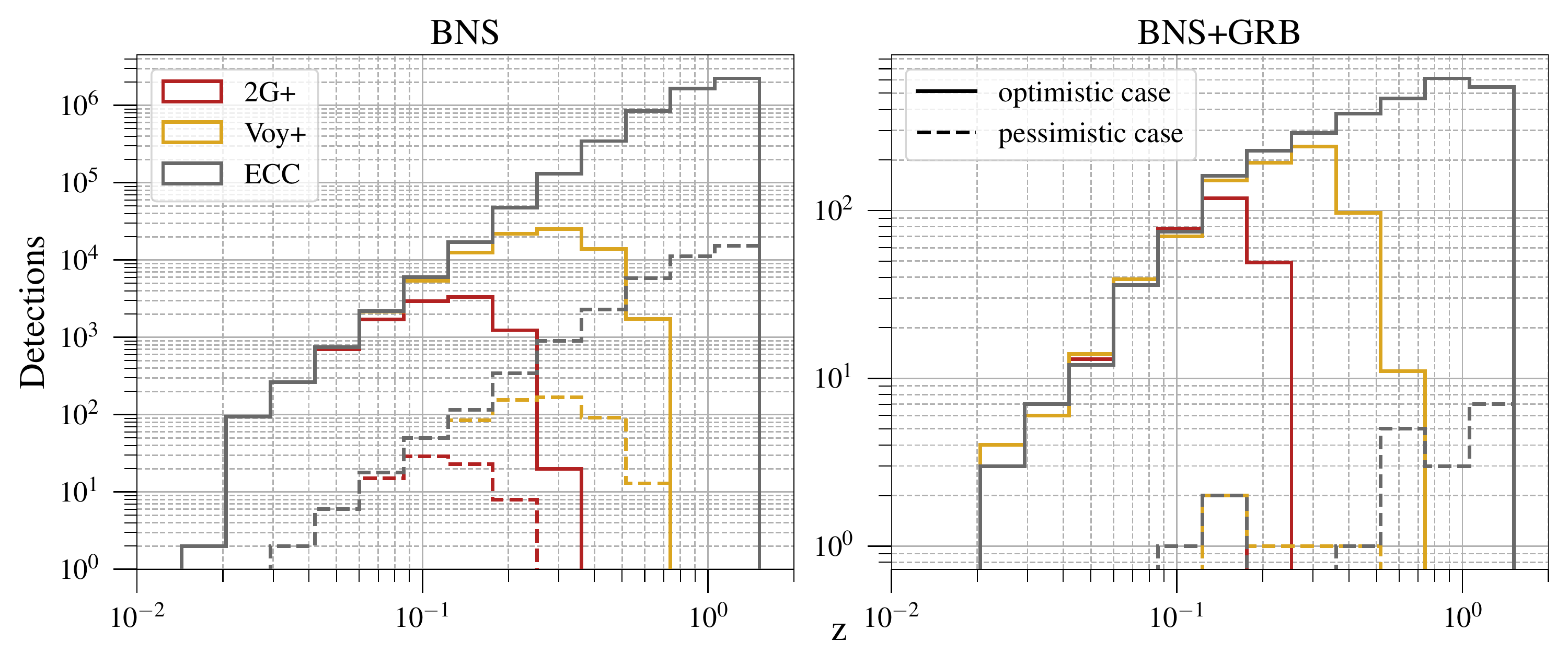}
    \caption{Number of GW (left panel) and GW+GRB (right panel) detections as a function of redshift for different detector networks considered in this study. The optimistic case (solid lines) represents the upper limit on the local BNS merger rate and the pessimistic case (dashed lines) the lower limit. The range of values for the local BNS merger rate is given in Eq.~(\ref{eq:local_merger_rate}). The lifetime of a network is assumed to be 10 years with an 80\% duty cycle for each detector.}
    \label{fig:grb_rates}
\end{figure*}

We calculate the rate of a coincident GRB detection following the procedure outlined in \textcite{Belgacem:2019tbw} and sketched it out here for completeness. We assume a Gaussian structured jet profile~\cite{Howell:2018nhu} for a GRB burst and the luminosity $L(\theta_V)$ is given by
\begin{equation}
    L(\theta_V) = L_p \exp\left(-\frac{\theta_V^2}{2 \theta_c^2}\right),
\end{equation}
where $\theta_V$ is the viewing angle and $\theta_c = 4.7\degree$ represents the variation in the GRB jet opening angle. $L_p$ is the peak luminosity of each burst assuming isotropic emission in the rest frame in the $1-10^4$ keV energy range and can be sampled from the probability distribution 
\begin{equation}
    \Phi(L_p) \propto
    \begin{cases}
        (L_p/L_*)^{\alpha},\qquad L_p<L_*, \\
        (L_p/L_*)^{\beta}, \qquad L_p\geq L_*,
\end{cases}
\end{equation}
where the parameters of the broken power-law distribution are $L_{*}=2 \times 10^{52} \,\rm erg/s$, $\alpha=-1.95$, and $\beta=-3$~\cite{Wanderman:2014eza}. A GRB is assumed to be detected if the observed peak flux $F_P(\theta_V) = L(\theta_V)/4\pi d_L^2$, given the GW luminosity distance and inclination angle, is greater than the flux limit of $1.1 \rm \,ph\, s^{-1}\, cm^{-2}$~\cite{Belgacem:2019tbw} in the $50$--300 keV band for Fermi-GBM. The total time-averaged observable sky fraction for the Fermi-GBM is taken to be 0.6~\cite{Burns:2015fol}.

\subsubsection{Rates for spatially coincident SNeIa}
\label{subsubsec:sne_rates}
Following the arguments presented in Sec. 3 of \textcite{Gupta:2019okl}, we now estimate the rates for a spatial coincidence of SNeIa and BNSs in a galaxy cluster. \textcite{Gupta:2019okl} concluded that the rate of spatial coincidence of SNeIa and BNS mergers in a galaxy given their rates \cite{2011MNRAS.412.1473L, LIGOScientific:2018mvr} is extremely small.
Moreover, in their Sec. 5 \textcite{Gupta:2019okl} showed that there is
${\cal O}(1\%)$ error in the distance estimation of SNeIa if calibrated through a BNS in the same galaxy cluster instead of the same galaxy. Therefore, coincident of SNeIa with a BNS in the same galaxy cluster is sufficient to obtain the redshift information of BNSs.

The current volumetric merger rate of BNSs is $13$--1900 $\rm Gpc^{-3}yr^{-1} $ \cite{LIGOScientific:2021psn} and that of local SNeIa is $3.0^{+0.6}_{-0.6}\times10^4 \, \rm Gpc^{-3}yr^{-1} $ \cite{2011MNRAS.412.1473L}. Considering the median of local SNeIa rates, it implies that there will be roughly 15 to 2300 SNeIa for a BNS merger in a galaxy. As in \cite{Gupta:2019okl}, we assume that the ratio of the SNeIa and BNS rates will be similar in rich galaxy clusters as well since both types of populations involve compact object mergers. For sources up to the redshift of $z=1.5$, we use SNeIa rate to be $0.65^{+0.61}_{-0.49}\times10^{-12}\, \rm L_{B,\odot}^{-1}yr^{-1}$  in rich galaxy clusters \cite{2018MNRAS.479.3563F}, where $L_\odot$ is the bolometric luminosity in solar units. Consequently, these numbers suggest that every year there will be $\approx 3$ SNeIa in a Coma-like cluster with bolometric luminosity of $L_{B} \approx 5.0\times10^{12}L_{B,\odot}$ \cite{Girardi:2002mmg} which is sufficient to  confirm the association with BNSs and derive their redshifts.


\subsection{Luminosity distance errors}
\label{subsec:lum_dist_err}
\subsubsection{Errors from gravitational-wave observations}
We simulate a population of neutron star binaries using the procedure outlined in Sec.~\ref{subsec:rates}. The redshift of a source is converted to its luminosity distance, the GW observable, using \textit{Planck18}~\cite{Aghanim:2018eyx} cosmology. For a BNS merger to be detectable, we require a network signal-to-noise ratio (SNR) threshold of 12 for each binary in the population but do not demand a minimum SNR for individual detectors. Note that the probability of having just one detector online in a 5 detector network with a duty cycle of 80\% for each detector is less than a percent~\cite{Belgacem:2019tbw}. We calculate the errors in the estimation of the binary parameters using the publicly available code, \textsc{gwbench}~\cite{Borhanian:2020ypi}, which implements a Fisher-matrix formalism~\cite{Cutler:1994ys,Poisson:1995ef} for error calculation. We use the \textsc{IMRPhenomPv2\_NRTidal} waveform model in our Fisher analysis, with a fixed effective tidal parameter $\Tilde{\Lambda}=100$. We do not compute the error on the  $\Tilde{\Lambda}$ measurement because this parameter is not expected to appreciably affect the luminosity distance estimate. We assume that the electromagnetic counterpart accurately provides the sky location, so we do not compute an error on it. We further take the chirp mass to be given because it is well estimated and mostly not degenerate with the luminosity distance. 
\begin{table*}[ht]
    \centering
    \begin{tabular}{c|c|c|c|c|c}
    \hline \hline
        Network & GW events & GW events ($z<0.5$) & GW + GRB events & GW + GRB events ($z>0.5$) & GW + EM counterpart\\
        \hline \hline
        2G+  & 10,259 (83)  & 10,259 (83) & 304 (2) & 0 (0) & 1330 (11) \\
        Voy+ & 83,697 (589) & 81,415 (571)  & 825 (5) & 17 (0) & 8,967 (62) \\
        ECC  & 5,286,423 (36,001) & 505,073 (3454) & 2,810 (19) & 1,657 (15) & 53,317 (364) \\
        \hline \hline
    \end{tabular}
    \caption{Number of GW events detected by the three networks in 10 years, together with the coincident GRB detection rate and the same for sources with $z>0.5$, assuming the detector characteristics of Fermi-GBM.}
    \label{tab:grb_rates}
\end{table*}
This slightly underestimates the distance errors but it would not affect our results significantly. From a technical perspective, this renders some of the otherwise ill-conditioned Fisher matrices of the 3G network to behave well. We are then left with a seven dimensional Fisher matrix consisting of the symmetric mass ratio $\eta$, the luminosity distance $d_L^{\rm GW}$, the inclination angle $\iota$, the polarization angle $\psi$, the time of coalescence $t_c$, and the phase of coalescence $\phi_c$. We subsequently extract the errors in the measurement of the luminosity distance which is the parameter of interest here.

Figure~\ref{fig:grb_rates} shows the redshift distribution of the detected GW events in our population (left panel), together with those that have an observable GRB counterpart in the Fermi-GBM detector (right panel) for the three different networks considered. The distribution is shown for both the optimistic case (solid lines) and the pessimistic case (dashed lines) corresponding to the range of the local BNS merger rates given in Eq.~(\ref{eq:local_merger_rate}). We note that only the 3G network can observe BNS coalescences from the furthest redshifts considered.
In Tab.~\ref{tab:grb_rates}, we quote the figures for the expected number of GW events, the corresponding number whose redshift is less than 0.5, the total number of events with a GRB counterpart, the number of events with a GRB counterpart above redshift $z>0.5$, and the cumulative number of events expected to contribute to the measurement of $\eta_0$ according to our assumptions in Sec.~\ref{sec:rate_em}. The numbers in parenthesis correspond to the pessimistic case. 

We see that the \textit{2G+} network has a horizon reach of less than $z<0.3$ and the total number of coincident electromagnetic detections  for the optimistic case are $\sim1330$ (304 + 10\% of 10,259). The corresponding numbers for the \textit{Voy+} and \textit{ECC} networks are $\sim8,970$ and $\sim53,320$, respectively, as given in the last column of Tab.~\ref{tab:grb_rates}.

\begin{figure}[h]
    \centering
    \includegraphics[width=\columnwidth]{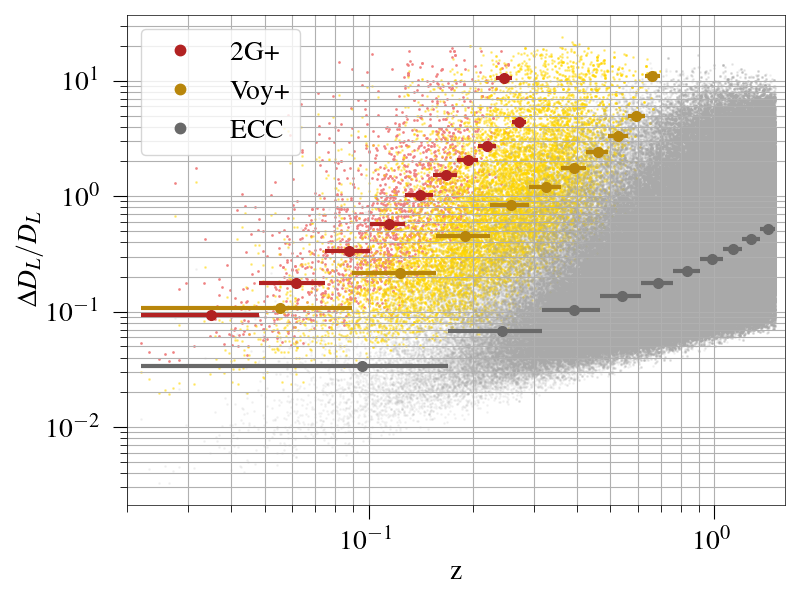}
    \caption{Fractional error in the measurement of gravitational luminosity distance as a function of redshift for simulated the BNS population. We see that only the third generation detector network detects sources from the highest redshift ($z=1.5$) considered in this study.}
    \label{fig:gw_dl_err}
\end{figure}

In Fig.~\ref{fig:gw_dl_err}, we show the fractional error in the measurement of GW luminosity distance as a function of redshift for the three detector networks for our detected population. To get the average behavior, we distribute the sources into redshift bins and calculate the median of the fractional errors of the sources in each redshift bin. We model the fractional luminosity distance errors as a function of redshift as a series of Heaviside step functions, which entails taking the errors in each redshift bin to be a constant.

We note that a fit for the fractional error in luminosity distance as a function of redshift can be found in~\textcite{Belgacem:2019tbw}. The reason we do not directly use their fits in our study is because the only parameter in their Fisher matrix is the luminosity distance and, therefore, their errors are unrealistic. Crucially, they ignore the correlations between the luminosity distance and inclination angle, which is known to increase the errors significantly~\cite{1993PhRvD..48.4738M,Cutler:1994ys}.

As can be seen from Fig.~\ref{fig:gw_dl_err}, the horizon distance for the \textit{2G+} network is $z\sim 0.3$ and hence we consider the full detectable population to have a possible kilonova counterpart detection. Another point of note from the figure is that the largest redshift considered in this study ($z=1.5$) is within the horizon distance for the \textit{ECC} network. We do not consider higher redshift sources because we are limited by the farthest observed SNeIa in the \textit{Union2} data-set (see Sec.~\ref{subsubsec:em_err}).

\subsubsection{Errors from electromagnetic observations}
\label{subsubsec:em_err}
\begin{figure}[h]
    \centering
    \includegraphics[width=\columnwidth]{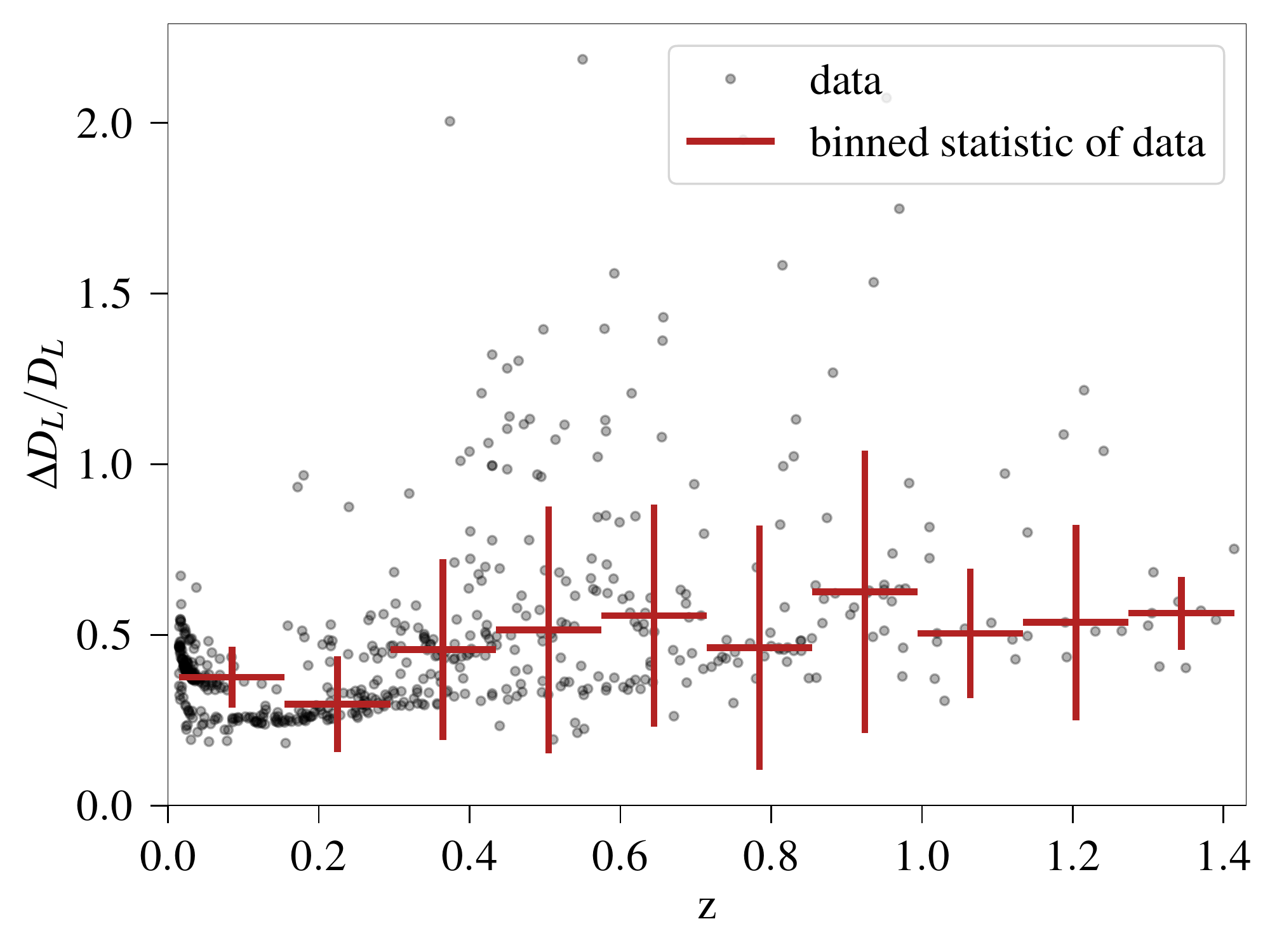}
    \caption{Fractional errors on SNe luminosity distances as a function of redshift. The red steps denote the median error for the corresponding redshift bin.}
    \label{fig:sne_err}
\end{figure}

We model the EM luminosity distance errors using the \textit{Union2}~\cite{2010ApJ...716..712A} SNeIa compilation. Supernovae distances are measured in units of distance modulus $\mu,$ which is related to luminosity distance by
\begin{equation}
    \mu = 5\log_{10}\left(\frac{10^5 d_L}{Mpc}\right).
\end{equation}
It can be easily seen from the above expression that the fractional error in the luminosity distance is given by
\begin{equation}
    \frac{\Delta d_L}{d_L} = \frac{5}{\log_e(10)} \Delta\mu.
\end{equation}
Figure~\ref{fig:sne_err} shows the fractional errors in the EM luminosity distance in the \textit{Union2} data-set as a function of redshift. We do not see any functional behaviour in the luminosity distance errors across redshift bins (with the errors approximately constant across bins) and, therefore, do not attempt at a fit, instead treating the redshift behaviour of the errors as a piece-wise step function. 

We also note that for the low redshift events of \textit{2G+} and \textit{Voy+} networks, we are limited by the SNeIa luminosity distance errors and the median errors for the \textit{ECC} network is always less than their SNeIa counterpart except around the redshift limit of $z=1.5$. 

The median errors in the distance modulus for the \textit{Union2} data-set is 0.19. The Rubin Observatory Legacy Survey of Space and Time is expected to observe about half a million supernovae in its survey life cycle of 10 years with a large fraction of them expected to have distance modulus errors of order 0.12 which is a $\sim40\%$ improvement over the \textit{Union2} data-set \cite{LSST:SNeIa}, which would further improve our estimates. 

\section{Results}
\label{sec:results}

\begin{figure}[h]
    \centering
    \includegraphics[width=\columnwidth]{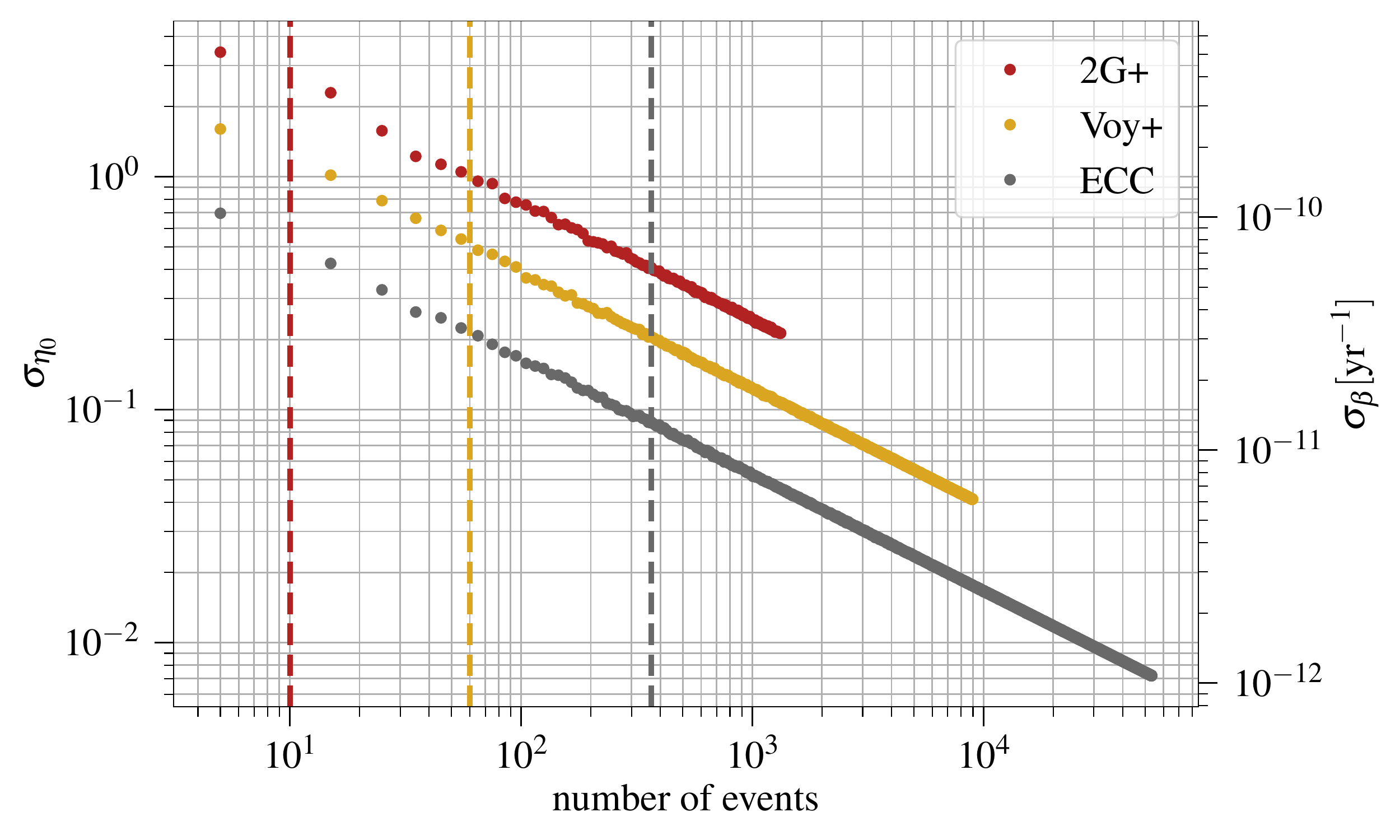}
    \caption{The expected error on the measurement of $\eta_0$ ($\sigma_{\eta_0}$) as a function of the number of observations for the three detector networks examined in this study. The axis on the right enumerates the same errors in terms of the temporal variation of the ratio of the gravitational and fine structure constant at the present epoch, $\beta$. The maximum number of events for each network denotes the expected number of total observations for the optimistic case. The corresponding number for the pessimistic case (rounded to the nearest multiple of 5) is shown by the vertical dashed lines.}
    \label{fig:err_eta}
\end{figure}

We calculate the errors on $\eta_0$ for our simulation as follows. From our sub-population of observed sources of GWs and their EM counterparts, we randomly select $N$ binaries. Given that we know the redshift to each of our sources, we get the median fractional error in the GW and EM luminosity distances for each of the $N$ detections from our modeling of the same as described in Sec.~\ref{subsec:lum_dist_err}. Assuming that the central value for both the luminosity distances are the same, we use Eq.~(\ref{eq:eta_err}) to calculate the errors on $\eta_0$ for each source. The combined error for $N$ independent observations is given by
\begin{equation}
    \frac{1}{\sigma^2} = \sum_{i=1}^N \frac{1}{\sigma_i^2},
\end{equation}
where $\sigma_i$ is the error for each event. We show the resultant errors on $\eta_0$ in Fig.~\ref{fig:err_eta} as a function of the number of observed events, in increments of 5 up to the expected number of observations for the optimistic case, for the three networks under consideration. The dashed vertical lines show the expected number of observations for the pessimistic case rounded to the nearest multiple of 5 for easy reading of the associated error. Also depicted in the figure on the right axis is the same error converted to the temporal variation of the ratio of the gravitational and fine structure constant at the current epoch [see Eq.~(\ref{eq:temp_var})] where $H_0=73.04\pm1.04\,\rm km/s/Mpc$~\cite{Riess:2021jrx}. We fit the $1/\sqrt{N}$ asymptotic behavior of the errors and quote the typical error for an observation in each network in Table~\ref{tab:err_eta}.

\begin{table}[h]
    \centering
    \begin{tabular}{ c|c c c|c }
        \hline \hline
        Network && $\sigma_{\eta_0}$ && $\sigma_{\beta}$ $[\times 10^{-9}\rm yr^{-1}]$ \\
        \hline 
        2G+ && 7.3 && 1.1 \\
        Voy+ && 4.1 && 0.61 \\
        ECC && 1.6 && 0.24 \\
        \hline \hline
    \end{tabular}
    \caption{Typical value of the error in the measurement of $\eta_0$ and the same in terms of the temporal variation of $G/\alpha(z=0)$ for a single GW detection for the three networks considered in this study.}
    \label{tab:err_eta}
\end{table}

We note that the constraints from the violation of the distance-duality relation directly translates to constraints on the variation of fundamental constants. We now compare our results to complimentary EM experiments that look for the variation of the fine structure constant from cosmological data. 
We stress that this comparison can only be done for a restricted class of models that do not modify the gravitational sector as the relevant EM experiments are oblivious to these modifications. 
\textcite{Hees:2014lfa} briefly reviewed such EM probes and the constraints from various probes are quoted in Table~\rom{1} of their paper. \textcite{Holanda:2012at} and \textcite{2011RAA....11.1199C} use the same parameterization of $\eta(z)$ as ours and quote average $1\sigma$ errors of 0.12 and 0.22 on $\eta_0$, respectively. The latter is on par with the capability of the \textit{2G+} network at the end of its observing cycle in the optimistic case. 

Other studies use different parameterizations but one can deduce that although the constraints using \textit{2G+} network's forecast to be of the same order or slightly more than the electromagnetic ones, \textit{Voy+} network would be able to place limits that are a few times better than most of the EM experiments---except constraints from high redshift quasar absorption spectra~\cite{Webb:2000mn,King:2012id,Webb:2010hc}---for the optimistic case. The \textit{ECC} network improves the \textit{Voy+} network constraints by an additional factor of 5. The pessimistic case yields constraints that are an order of magnitude poorer than the optimistic case for all the three networks studied here, which is in line with the $1/\sqrt{N}$ behaviour of errors -- the optimistic case has two orders of magnitude more events than the pessimistic case. Furthermore, observations of quasars from high redshifts suggest a possible spatial variation of the fine structure constant~\cite{King:2012id,Webb:2010hc}. The large number of gravitational wave observations, albeit from smaller redshifts, would allow for local constraints on the spatial variation of the fine structure constant too.

\section{Constraints based on GW170817}
\label{sec:gw170817_constraints}
In the previous sections, we focused on the detection of a spatially coincident SNeIa to provide the EM luminosity distance. This is because a SNeIa is a standard candle and, hence, has constant absolute luminosity in the source frame. In addition, the systematic uncertainties of modeling SNeIa as a standard candles is well understood and, therefore, provides an unbiased estimate for the luminosity distance to the source.

Recently, there have been efforts to model the kilonova emissions following a BNS merger as a standard candle~\cite{Kashyap:2019ypm,Coughlin:2019vtv}. This would provide another independent measure of the EM luminosity distance for low redshift sources with the added benefit of not having to search for a spatially coincident supernova. We, however, do not forecast the constraints that can be placed on $\eta_0$ for a population of joint GW-kilonova sources using standardised kilonova emissions since these models are at a very nascent stage of development with large systematic uncertainties.
Nevertheless, we use the EM luminosity distance estimates of \textcite{Coughlin:2019vtv} for GW170817~\cite{LIGOScientific:2017vwq} to place the first constraints on our deviation parameter $\eta_0$. 
The gravitational-wave luminosity distance for GW170817 was estimated to be $d_L^{\rm GW} = 43.8^{+2.9}_{-6.9}\,\text{Mpc}$~\cite{Abbott:2017xzu}. \textcite{Coughlin:2019vtv} give three measurements of the EM luminosity distance. The first value, $d_L^{\rm EM}=31^{+17}_{-11}\,\text{Mpc}$, is a direct measurement from the lightcurve based on the analysis of~\textcite{Kasen:2017sxr}. The other two are inferred from ejecta parameters based on the analyses of \textcite{Kasen:2017sxr} and \textcite{Bulla:2019muo} and are given by $d_L^{\rm EM}=37^{+8}_{-7}\,\text{Mpc}$ and $d_L^{\rm EM}=40^{+9}_{-8}\,\text{Mpc}$, respectively. These three luminosity distance measurements correspond to $\eta_0=42^{+79}_{-55}$, $18^{+27}_{-29}$, and $10^{+26}_{-28}$, respectively. Unsurprisingly, the estimate of $\eta_0$ is consistent with 0. 

The above measurements of $\eta_0$ give the temporal variation of $G/\alpha(z)$ [see Eq.~(\ref{eq:temp_var})] -- in units of $[\times10^{-9} \rm yr^{-1}]$ -- at the current epoch to be 
\begin{equation}
    \beta = -6^{+8}_{-12}, \; -3^{+4}_{-4}, \; -1^{+4}_{-4},
\end{equation}
respectively. A Hubble constant value of $H_0=73.04\pm1.04\, \rm km/s/Mpc$ as reported by~\textcite{Riess:2021jrx} is used for this calculation.

\section{Conclusion}
\label{sec:conclusion}
In this paper, we focused on constraining the ratio of gravitational-wave and electromagnetic-wave luminosity distances and, consequently, the variation in the ratio of the gravitational and fine structure constant, using coincident gravitational and electromagnetic wave observations for a class of scalar-tensor theories known as {\it runaway dilaton} models. These theories have multiplicative couplings of generic scalar fields to gravitational and electromagnetic sectors. To constrain the modified propagation in such theories without having to fit for other cosmological parameters, as is done while using a luminosity distance-redshift relation, a second distance measure is necessary. We used a spatially coincident supernova as the EM probe to provide the complimentary EM luminosity distance estimate. 

We find that the planned upgrade to the current second-generation ground-based detector network (\textit{2G+}) can constrain the parameter modeling the ratio of the EM and GW luminosity distance to below $|\eta_0| \lesssim 0.2,$ while the proposed improvement of the \textit{2G+} network to \textit{Voy+} sensitivity would be able to place an upper limit of $|\eta_0| \lesssim 0.05$, at the end of an 8-year effective observing cycle if no deviation from the GR value of $\eta_0 = 0$ is measured. The proposed next-generation ground-based detector network (\textit{ECC}) can further improve the constraints to $|\eta_0| \lesssim 0.01$.
We see that the constraints using this method for the sub-class of theories that modify the EM sector alone would be competitive with most of the current EM probes in the literature~\cite{Hees:2014lfa} for the \textit{2G+} network. The \textit{Voy+} network would improve these estimates by a factor of 4 and the \textit{ECC} network improves the \textit{Voy+} network constraints by a factor of 5. We also showed how these numbers translate in to the temporal variation of the fundamental constants. 

We, further, make use of recent progress in kilonova light-curve modeling and, consequently, the EM luminosity distance estimates from them to place the first constraints on our $\eta_0$ parameter for GW170817. As expected, we find consistency with GR.

We expect a number of BNS merger observations with counterpart in the fourth observing run of aLIGO/aVirgo/KAGRA and plan to use them to constrain this class of theories.


\begin{acknowledgements}
AD was supported by NSF grant PHY-2012083 and BSS was supported in part by NSF grants PHY-1836779, PHY-2012083 and AST-2006384. 
\end{acknowledgements}

\bibliography{alpha_z.bib}{}

\begin{thebibliography}{111}%
\makeatletter
\providecommand \@ifxundefined [1]{%
 \@ifx{#1\undefined}
}%
\providecommand \@ifnum [1]{%
 \ifnum #1\expandafter \@firstoftwo
 \else \expandafter \@secondoftwo
 \fi
}%
\providecommand \@ifx [1]{%
 \ifx #1\expandafter \@firstoftwo
 \else \expandafter \@secondoftwo
 \fi
}%
\providecommand \natexlab [1]{#1}%
\providecommand \enquote  [1]{``#1''}%
\providecommand \bibnamefont  [1]{#1}%
\providecommand \bibfnamefont [1]{#1}%
\providecommand \citenamefont [1]{#1}%
\providecommand \href@noop [0]{\@secondoftwo}%
\providecommand \href [0]{\begingroup \@sanitize@url \@href}%
\providecommand \@href[1]{\@@startlink{#1}\@@href}%
\providecommand \@@href[1]{\endgroup#1\@@endlink}%
\providecommand \@sanitize@url [0]{\catcode `\\12\catcode `\$12\catcode
  `\&12\catcode `\#12\catcode `\^12\catcode `\_12\catcode `\%12\relax}%
\providecommand \@@startlink[1]{}%
\providecommand \@@endlink[0]{}%
\providecommand \url  [0]{\begingroup\@sanitize@url \@url }%
\providecommand \@url [1]{\endgroup\@href {#1}{\urlprefix }}%
\providecommand \urlprefix  [0]{URL }%
\providecommand \Eprint [0]{\href }%
\providecommand \doibase [0]{http://dx.doi.org/}%
\providecommand \selectlanguage [0]{\@gobble}%
\providecommand \bibinfo  [0]{\@secondoftwo}%
\providecommand \bibfield  [0]{\@secondoftwo}%
\providecommand \translation [1]{[#1]}%
\providecommand \BibitemOpen [0]{}%
\providecommand \bibitemStop [0]{}%
\providecommand \bibitemNoStop [0]{.\EOS\space}%
\providecommand \EOS [0]{\spacefactor3000\relax}%
\providecommand \BibitemShut  [1]{\csname bibitem#1\endcsname}%
\let\auto@bib@innerbib\@empty
\bibitem [{\citenamefont {Maggiore}(2007)}]{Maggiore:1900zz}%
  \BibitemOpen
  \bibfield  {author} {\bibinfo {author} {\bibfnamefont {M.}~\bibnamefont
  {Maggiore}},\ }\href@noop {} {\emph {\bibinfo {title} {{Gravitational Waves.
  Vol. 1: Theory and Experiments}}}},\ Oxford Master Series in Physics\
  (\bibinfo  {publisher} {Oxford University Press},\ \bibinfo {year}
  {2007})\BibitemShut {NoStop}%
\bibitem [{\citenamefont {Fujii}\ and\ \citenamefont
  {Maeda}(2007)}]{Fujii:2003pa}%
  \BibitemOpen
  \bibfield  {author} {\bibinfo {author} {\bibfnamefont {Y.}~\bibnamefont
  {Fujii}}\ and\ \bibinfo {author} {\bibfnamefont {K.}~\bibnamefont {Maeda}},\
  }\href {\doibase 10.1017/CBO9780511535093} {\emph {\bibinfo {title} {{The
  scalar-tensor theory of gravitation}}}},\ Cambridge Monographs on
  Mathematical Physics\ (\bibinfo  {publisher} {Cambridge University Press},\
  \bibinfo {year} {2007})\BibitemShut {NoStop}%
\bibitem [{\citenamefont {Bertolami}\ \emph {et~al.}(2007)\citenamefont
  {Bertolami}, \citenamefont {Boehmer}, \citenamefont {Harko},\ and\
  \citenamefont {Lobo}}]{Bertolami:2007gv}%
  \BibitemOpen
  \bibfield  {author} {\bibinfo {author} {\bibfnamefont {O.}~\bibnamefont
  {Bertolami}}, \bibinfo {author} {\bibfnamefont {C.~G.}\ \bibnamefont
  {Boehmer}}, \bibinfo {author} {\bibfnamefont {T.}~\bibnamefont {Harko}}, \
  and\ \bibinfo {author} {\bibfnamefont {F.~S.~N.}\ \bibnamefont {Lobo}},\
  }\href {\doibase 10.1103/PhysRevD.75.104016} {\bibfield  {journal} {\bibinfo
  {journal} {Phys. Rev. D}\ }\textbf {\bibinfo {volume} {75}},\ \bibinfo
  {pages} {104016} (\bibinfo {year} {2007})},\ \Eprint
  {http://arxiv.org/abs/0704.1733} {arXiv:0704.1733 [gr-qc]} \BibitemShut
  {NoStop}%
\bibitem [{\citenamefont {Bertolami}\ \emph {et~al.}(2008)\citenamefont
  {Bertolami}, \citenamefont {Lobo},\ and\ \citenamefont
  {Paramos}}]{Bertolami:2008ab}%
  \BibitemOpen
  \bibfield  {author} {\bibinfo {author} {\bibfnamefont {O.}~\bibnamefont
  {Bertolami}}, \bibinfo {author} {\bibfnamefont {F.~S.~N.}\ \bibnamefont
  {Lobo}}, \ and\ \bibinfo {author} {\bibfnamefont {J.}~\bibnamefont
  {Paramos}},\ }\href {\doibase 10.1103/PhysRevD.78.064036} {\bibfield
  {journal} {\bibinfo  {journal} {Phys. Rev. D}\ }\textbf {\bibinfo {volume}
  {78}},\ \bibinfo {pages} {064036} (\bibinfo {year} {2008})},\ \Eprint
  {http://arxiv.org/abs/0806.4434} {arXiv:0806.4434 [gr-qc]} \BibitemShut
  {NoStop}%
\bibitem [{\citenamefont {Bertolami}\ and\ \citenamefont
  {Paramos}(2008)}]{Bertolami:2008im}%
  \BibitemOpen
  \bibfield  {author} {\bibinfo {author} {\bibfnamefont {O.}~\bibnamefont
  {Bertolami}}\ and\ \bibinfo {author} {\bibfnamefont {J.}~\bibnamefont
  {Paramos}},\ }\href {\doibase 10.1088/0264-9381/25/24/245017} {\bibfield
  {journal} {\bibinfo  {journal} {Class. Quant. Grav.}\ }\textbf {\bibinfo
  {volume} {25}},\ \bibinfo {pages} {245017} (\bibinfo {year} {2008})},\
  \Eprint {http://arxiv.org/abs/0805.1241} {arXiv:0805.1241 [gr-qc]}
  \BibitemShut {NoStop}%
\bibitem [{\citenamefont {Sotiriou}\ and\ \citenamefont
  {Faraoni}(2008)}]{Sotiriou:2008it}%
  \BibitemOpen
  \bibfield  {author} {\bibinfo {author} {\bibfnamefont {T.~P.}\ \bibnamefont
  {Sotiriou}}\ and\ \bibinfo {author} {\bibfnamefont {V.}~\bibnamefont
  {Faraoni}},\ }\href {\doibase 10.1088/0264-9381/25/20/205002} {\bibfield
  {journal} {\bibinfo  {journal} {Class. Quant. Grav.}\ }\textbf {\bibinfo
  {volume} {25}},\ \bibinfo {pages} {205002} (\bibinfo {year} {2008})},\
  \Eprint {http://arxiv.org/abs/0805.1249} {arXiv:0805.1249 [gr-qc]}
  \BibitemShut {NoStop}%
\bibitem [{\citenamefont {De~Felice}\ and\ \citenamefont
  {Tsujikawa}(2010)}]{DeFelice:2010aj}%
  \BibitemOpen
  \bibfield  {author} {\bibinfo {author} {\bibfnamefont {A.}~\bibnamefont
  {De~Felice}}\ and\ \bibinfo {author} {\bibfnamefont {S.}~\bibnamefont
  {Tsujikawa}},\ }\href {\doibase 10.12942/lrr-2010-3} {\bibfield  {journal}
  {\bibinfo  {journal} {Living Rev. Rel.}\ }\textbf {\bibinfo {volume} {13}},\
  \bibinfo {pages} {3} (\bibinfo {year} {2010})},\ \Eprint
  {http://arxiv.org/abs/1002.4928} {arXiv:1002.4928 [gr-qc]} \BibitemShut
  {NoStop}%
\bibitem [{\citenamefont {Harko}\ \emph {et~al.}(2013)\citenamefont {Harko},
  \citenamefont {Lobo},\ and\ \citenamefont {Minazzoli}}]{Harko:2012hm}%
  \BibitemOpen
  \bibfield  {author} {\bibinfo {author} {\bibfnamefont {T.}~\bibnamefont
  {Harko}}, \bibinfo {author} {\bibfnamefont {F.~S.~N.}\ \bibnamefont {Lobo}},
  \ and\ \bibinfo {author} {\bibfnamefont {O.}~\bibnamefont {Minazzoli}},\
  }\href {\doibase 10.1103/PhysRevD.87.047501} {\bibfield  {journal} {\bibinfo
  {journal} {Phys. Rev. D}\ }\textbf {\bibinfo {volume} {87}},\ \bibinfo
  {pages} {047501} (\bibinfo {year} {2013})},\ \Eprint
  {http://arxiv.org/abs/1210.4218} {arXiv:1210.4218 [gr-qc]} \BibitemShut
  {NoStop}%
\bibitem [{\citenamefont {Das}\ and\ \citenamefont
  {Banerjee}(2008)}]{Das:2008iq}%
  \BibitemOpen
  \bibfield  {author} {\bibinfo {author} {\bibfnamefont {S.}~\bibnamefont
  {Das}}\ and\ \bibinfo {author} {\bibfnamefont {N.}~\bibnamefont {Banerjee}},\
  }\href {\doibase 10.1103/PhysRevD.78.043512} {\bibfield  {journal} {\bibinfo
  {journal} {Phys. Rev. D}\ }\textbf {\bibinfo {volume} {78}},\ \bibinfo
  {pages} {043512} (\bibinfo {year} {2008})},\ \Eprint
  {http://arxiv.org/abs/0803.3936} {arXiv:0803.3936 [gr-qc]} \BibitemShut
  {NoStop}%
\bibitem [{\citenamefont {Bisabr}(2012)}]{Bisabr:2012cu}%
  \BibitemOpen
  \bibfield  {author} {\bibinfo {author} {\bibfnamefont {Y.}~\bibnamefont
  {Bisabr}},\ }\href {\doibase 10.1103/PhysRevD.86.127503} {\bibfield
  {journal} {\bibinfo  {journal} {Phys. Rev. D}\ }\textbf {\bibinfo {volume}
  {86}},\ \bibinfo {pages} {127503} (\bibinfo {year} {2012})},\ \Eprint
  {http://arxiv.org/abs/1212.2709} {arXiv:1212.2709 [gr-qc]} \BibitemShut
  {NoStop}%
\bibitem [{\citenamefont {Moffat}\ and\ \citenamefont
  {Toth}(2012)}]{Moffat:2010ek}%
  \BibitemOpen
  \bibfield  {author} {\bibinfo {author} {\bibfnamefont {J.~W.}\ \bibnamefont
  {Moffat}}\ and\ \bibinfo {author} {\bibfnamefont {V.~T.}\ \bibnamefont
  {Toth}},\ }\href {\doibase 10.1142/S0218271812500848} {\bibfield  {journal}
  {\bibinfo  {journal} {Int. J. Mod. Phys. D}\ }\textbf {\bibinfo {volume}
  {21}},\ \bibinfo {pages} {1250084} (\bibinfo {year} {2012})},\ \Eprint
  {http://arxiv.org/abs/1001.1564} {arXiv:1001.1564 [gr-qc]} \BibitemShut
  {NoStop}%
\bibitem [{\citenamefont {Shiralilou}\ \emph {et~al.}(2022)\citenamefont
  {Shiralilou}, \citenamefont {Hinderer}, \citenamefont {Nissanke},
  \citenamefont {Ortiz},\ and\ \citenamefont {Witek}}]{Shiralilou:2021mfl}%
  \BibitemOpen
  \bibfield  {author} {\bibinfo {author} {\bibfnamefont {B.}~\bibnamefont
  {Shiralilou}}, \bibinfo {author} {\bibfnamefont {T.}~\bibnamefont
  {Hinderer}}, \bibinfo {author} {\bibfnamefont {S.~M.}\ \bibnamefont
  {Nissanke}}, \bibinfo {author} {\bibfnamefont {N.}~\bibnamefont {Ortiz}}, \
  and\ \bibinfo {author} {\bibfnamefont {H.}~\bibnamefont {Witek}},\ }\href
  {\doibase 10.1088/1361-6382/ac4196} {\bibfield  {journal} {\bibinfo
  {journal} {Class. Quant. Grav.}\ }\textbf {\bibinfo {volume} {39}},\ \bibinfo
  {pages} {035002} (\bibinfo {year} {2022})},\ \Eprint
  {http://arxiv.org/abs/2105.13972} {arXiv:2105.13972 [gr-qc]} \BibitemShut
  {NoStop}%
\bibitem [{\citenamefont {Rovelli}\ and\ \citenamefont
  {Smolin}(1994)}]{Rovelli:1993bm}%
  \BibitemOpen
  \bibfield  {author} {\bibinfo {author} {\bibfnamefont {C.}~\bibnamefont
  {Rovelli}}\ and\ \bibinfo {author} {\bibfnamefont {L.}~\bibnamefont
  {Smolin}},\ }\href {\doibase 10.1103/PhysRevLett.72.446} {\bibfield
  {journal} {\bibinfo  {journal} {Phys. Rev. Lett.}\ }\textbf {\bibinfo
  {volume} {72}},\ \bibinfo {pages} {446} (\bibinfo {year} {1994})},\ \Eprint
  {http://arxiv.org/abs/gr-qc/9308002} {arXiv:gr-qc/9308002} \BibitemShut
  {NoStop}%
\bibitem [{\citenamefont {Domagala}\ \emph {et~al.}(2010)\citenamefont
  {Domagala}, \citenamefont {Giesel}, \citenamefont {Kaminski},\ and\
  \citenamefont {Lewandowski}}]{Domagala:2010bm}%
  \BibitemOpen
  \bibfield  {author} {\bibinfo {author} {\bibfnamefont {M.}~\bibnamefont
  {Domagala}}, \bibinfo {author} {\bibfnamefont {K.}~\bibnamefont {Giesel}},
  \bibinfo {author} {\bibfnamefont {W.}~\bibnamefont {Kaminski}}, \ and\
  \bibinfo {author} {\bibfnamefont {J.}~\bibnamefont {Lewandowski}},\ }\href
  {\doibase 10.1103/PhysRevD.82.104038} {\bibfield  {journal} {\bibinfo
  {journal} {Phys. Rev. D}\ }\textbf {\bibinfo {volume} {82}},\ \bibinfo
  {pages} {104038} (\bibinfo {year} {2010})},\ \Eprint
  {http://arxiv.org/abs/1009.2445} {arXiv:1009.2445 [gr-qc]} \BibitemShut
  {NoStop}%
\bibitem [{\citenamefont {Green}\ \emph {et~al.}(1988)\citenamefont {Green},
  \citenamefont {Schwarz},\ and\ \citenamefont {Witten}}]{Green:1987mn}%
  \BibitemOpen
  \bibfield  {author} {\bibinfo {author} {\bibfnamefont {M.~B.}\ \bibnamefont
  {Green}}, \bibinfo {author} {\bibfnamefont {J.~H.}\ \bibnamefont {Schwarz}},
  \ and\ \bibinfo {author} {\bibfnamefont {E.}~\bibnamefont {Witten}},\
  }\href@noop {} {\emph {\bibinfo {title} {{SUPERSTRING THEORY. VOL. 2: LOOP
  AMPLITUDES, ANOMALIES AND PHENOMENOLOGY}}}}\ (\bibinfo  {publisher}
  {Cambridge University Press},\ \bibinfo {year} {1988})\BibitemShut {NoStop}%
\bibitem [{\citenamefont {Uzan}(2011)}]{Uzan:2010pm}%
  \BibitemOpen
  \bibfield  {author} {\bibinfo {author} {\bibfnamefont {J.-P.}\ \bibnamefont
  {Uzan}},\ }\href {\doibase 10.12942/lrr-2011-2} {\bibfield  {journal}
  {\bibinfo  {journal} {Living Rev. Rel.}\ }\textbf {\bibinfo {volume} {14}},\
  \bibinfo {pages} {2} (\bibinfo {year} {2011})},\ \Eprint
  {http://arxiv.org/abs/1009.5514} {arXiv:1009.5514 [astro-ph.CO]} \BibitemShut
  {NoStop}%
\bibitem [{\citenamefont {Damour}\ and\ \citenamefont
  {Polyakov}(1994{\natexlab{a}})}]{Damour:1994ya}%
  \BibitemOpen
  \bibfield  {author} {\bibinfo {author} {\bibfnamefont {T.}~\bibnamefont
  {Damour}}\ and\ \bibinfo {author} {\bibfnamefont {A.~M.}\ \bibnamefont
  {Polyakov}},\ }\href {\doibase 10.1007/BF02106709} {\bibfield  {journal}
  {\bibinfo  {journal} {Gen. Rel. Grav.}\ }\textbf {\bibinfo {volume} {26}},\
  \bibinfo {pages} {1171} (\bibinfo {year} {1994}{\natexlab{a}})},\ \Eprint
  {http://arxiv.org/abs/gr-qc/9411069} {arXiv:gr-qc/9411069} \BibitemShut
  {NoStop}%
\bibitem [{\citenamefont {Damour}\ and\ \citenamefont
  {Polyakov}(1994{\natexlab{b}})}]{Damour:1994zq}%
  \BibitemOpen
  \bibfield  {author} {\bibinfo {author} {\bibfnamefont {T.}~\bibnamefont
  {Damour}}\ and\ \bibinfo {author} {\bibfnamefont {A.~M.}\ \bibnamefont
  {Polyakov}},\ }\href {\doibase 10.1016/0550-3213(94)90143-0} {\bibfield
  {journal} {\bibinfo  {journal} {Nucl. Phys. B}\ }\textbf {\bibinfo {volume}
  {423}},\ \bibinfo {pages} {532} (\bibinfo {year} {1994}{\natexlab{b}})},\
  \Eprint {http://arxiv.org/abs/hep-th/9401069} {arXiv:hep-th/9401069}
  \BibitemShut {NoStop}%
\bibitem [{\citenamefont {Gasperini}\ \emph {et~al.}(2002)\citenamefont
  {Gasperini}, \citenamefont {Piazza},\ and\ \citenamefont
  {Veneziano}}]{Gasperini:2001pc}%
  \BibitemOpen
  \bibfield  {author} {\bibinfo {author} {\bibfnamefont {M.}~\bibnamefont
  {Gasperini}}, \bibinfo {author} {\bibfnamefont {F.}~\bibnamefont {Piazza}}, \
  and\ \bibinfo {author} {\bibfnamefont {G.}~\bibnamefont {Veneziano}},\ }\href
  {\doibase 10.1103/PhysRevD.65.023508} {\bibfield  {journal} {\bibinfo
  {journal} {Phys. Rev. D}\ }\textbf {\bibinfo {volume} {65}},\ \bibinfo
  {pages} {023508} (\bibinfo {year} {2002})},\ \Eprint
  {http://arxiv.org/abs/gr-qc/0108016} {arXiv:gr-qc/0108016} \BibitemShut
  {NoStop}%
\bibitem [{\citenamefont {Minazzoli}\ and\ \citenamefont
  {Hees}(2013)}]{Minazzoli:2013ara}%
  \BibitemOpen
  \bibfield  {author} {\bibinfo {author} {\bibfnamefont {O.}~\bibnamefont
  {Minazzoli}}\ and\ \bibinfo {author} {\bibfnamefont {A.}~\bibnamefont
  {Hees}},\ }\href {\doibase 10.1103/PhysRevD.88.041504} {\bibfield  {journal}
  {\bibinfo  {journal} {Phys. Rev. D}\ }\textbf {\bibinfo {volume} {88}},\
  \bibinfo {pages} {041504} (\bibinfo {year} {2013})},\ \Eprint
  {http://arxiv.org/abs/1308.2770} {arXiv:1308.2770 [gr-qc]} \BibitemShut
  {NoStop}%
\bibitem [{\citenamefont {Ratra}\ and\ \citenamefont
  {Peebles}(1988)}]{Ratra:1987rm}%
  \BibitemOpen
  \bibfield  {author} {\bibinfo {author} {\bibfnamefont {B.}~\bibnamefont
  {Ratra}}\ and\ \bibinfo {author} {\bibfnamefont {P.~J.~E.}\ \bibnamefont
  {Peebles}},\ }\href {\doibase 10.1103/PhysRevD.37.3406} {\bibfield  {journal}
  {\bibinfo  {journal} {Phys. Rev. D}\ }\textbf {\bibinfo {volume} {37}},\
  \bibinfo {pages} {3406} (\bibinfo {year} {1988})}\BibitemShut {NoStop}%
\bibitem [{\citenamefont {Caldwell}\ \emph {et~al.}(1998)\citenamefont
  {Caldwell}, \citenamefont {Dave},\ and\ \citenamefont
  {Steinhardt}}]{Caldwell:1997ii}%
  \BibitemOpen
  \bibfield  {author} {\bibinfo {author} {\bibfnamefont {R.~R.}\ \bibnamefont
  {Caldwell}}, \bibinfo {author} {\bibfnamefont {R.}~\bibnamefont {Dave}}, \
  and\ \bibinfo {author} {\bibfnamefont {P.~J.}\ \bibnamefont {Steinhardt}},\
  }\href {\doibase 10.1103/PhysRevLett.80.1582} {\bibfield  {journal} {\bibinfo
   {journal} {Phys. Rev. Lett.}\ }\textbf {\bibinfo {volume} {80}},\ \bibinfo
  {pages} {1582} (\bibinfo {year} {1998})},\ \Eprint
  {http://arxiv.org/abs/astro-ph/9708069} {arXiv:astro-ph/9708069} \BibitemShut
  {NoStop}%
\bibitem [{\citenamefont {Peebles}\ and\ \citenamefont
  {Ratra}(2003)}]{Peebles:2002gy}%
  \BibitemOpen
  \bibfield  {author} {\bibinfo {author} {\bibfnamefont {P.~J.~E.}\
  \bibnamefont {Peebles}}\ and\ \bibinfo {author} {\bibfnamefont
  {B.}~\bibnamefont {Ratra}},\ }\href {\doibase 10.1103/RevModPhys.75.559}
  {\bibfield  {journal} {\bibinfo  {journal} {Rev. Mod. Phys.}\ }\textbf
  {\bibinfo {volume} {75}},\ \bibinfo {pages} {559} (\bibinfo {year} {2003})},\
  \Eprint {http://arxiv.org/abs/astro-ph/0207347} {arXiv:astro-ph/0207347}
  \BibitemShut {NoStop}%
\bibitem [{\citenamefont {Guth}(1981)}]{Guth:1980zm}%
  \BibitemOpen
  \bibfield  {author} {\bibinfo {author} {\bibfnamefont {A.~H.}\ \bibnamefont
  {Guth}},\ }\href {\doibase 10.1103/PhysRevD.23.347} {\bibfield  {journal}
  {\bibinfo  {journal} {Phys. Rev. D}\ }\textbf {\bibinfo {volume} {23}},\
  \bibinfo {pages} {347} (\bibinfo {year} {1981})}\BibitemShut {NoStop}%
\bibitem [{\citenamefont {Linde}(1982)}]{Linde:1981mu}%
  \BibitemOpen
  \bibfield  {author} {\bibinfo {author} {\bibfnamefont {A.~D.}\ \bibnamefont
  {Linde}},\ }\href {\doibase 10.1016/0370-2693(82)91219-9} {\bibfield
  {journal} {\bibinfo  {journal} {Phys. Lett. B}\ }\textbf {\bibinfo {volume}
  {108}},\ \bibinfo {pages} {389} (\bibinfo {year} {1982})}\BibitemShut
  {NoStop}%
\bibitem [{\citenamefont {Albrecht}\ and\ \citenamefont
  {Steinhardt}(1982)}]{Albrecht:1982wi}%
  \BibitemOpen
  \bibfield  {author} {\bibinfo {author} {\bibfnamefont {A.}~\bibnamefont
  {Albrecht}}\ and\ \bibinfo {author} {\bibfnamefont {P.~J.}\ \bibnamefont
  {Steinhardt}},\ }\href {\doibase 10.1103/PhysRevLett.48.1220} {\bibfield
  {journal} {\bibinfo  {journal} {Phys. Rev. Lett.}\ }\textbf {\bibinfo
  {volume} {48}},\ \bibinfo {pages} {1220} (\bibinfo {year}
  {1982})}\BibitemShut {NoStop}%
\bibitem [{\citenamefont {Linde}(2008)}]{Linde:2007fr}%
  \BibitemOpen
  \bibfield  {author} {\bibinfo {author} {\bibfnamefont {A.~D.}\ \bibnamefont
  {Linde}},\ }\href {\doibase 10.1007/978-3-540-74353-8_1} {\bibfield
  {journal} {\bibinfo  {journal} {Lect. Notes Phys.}\ }\textbf {\bibinfo
  {volume} {738}},\ \bibinfo {pages} {1} (\bibinfo {year} {2008})},\ \Eprint
  {http://arxiv.org/abs/0705.0164} {arXiv:0705.0164 [hep-th]} \BibitemShut
  {NoStop}%
\bibitem [{\citenamefont {Bekenstein}(1982)}]{Bekenstein:1982eu}%
  \BibitemOpen
  \bibfield  {author} {\bibinfo {author} {\bibfnamefont {J.~D.}\ \bibnamefont
  {Bekenstein}},\ }\href {\doibase 10.1103/PhysRevD.25.1527} {\bibfield
  {journal} {\bibinfo  {journal} {Phys. Rev. D}\ }\textbf {\bibinfo {volume}
  {25}},\ \bibinfo {pages} {1527} (\bibinfo {year} {1982})}\BibitemShut
  {NoStop}%
\bibitem [{\citenamefont {Sandvik}\ \emph {et~al.}(2002)\citenamefont
  {Sandvik}, \citenamefont {Barrow},\ and\ \citenamefont
  {Magueijo}}]{Sandvik:2001rv}%
  \BibitemOpen
  \bibfield  {author} {\bibinfo {author} {\bibfnamefont {H.~B.}\ \bibnamefont
  {Sandvik}}, \bibinfo {author} {\bibfnamefont {J.~D.}\ \bibnamefont {Barrow}},
  \ and\ \bibinfo {author} {\bibfnamefont {J.}~\bibnamefont {Magueijo}},\
  }\href {\doibase 10.1103/PhysRevLett.88.031302} {\bibfield  {journal}
  {\bibinfo  {journal} {Phys. Rev. Lett.}\ }\textbf {\bibinfo {volume} {88}},\
  \bibinfo {pages} {031302} (\bibinfo {year} {2002})},\ \Eprint
  {http://arxiv.org/abs/astro-ph/0107512} {arXiv:astro-ph/0107512} \BibitemShut
  {NoStop}%
\bibitem [{\citenamefont {Dvali}\ and\ \citenamefont
  {Zaldarriaga}(2002)}]{Dvali:2001dd}%
  \BibitemOpen
  \bibfield  {author} {\bibinfo {author} {\bibfnamefont {G.~R.}\ \bibnamefont
  {Dvali}}\ and\ \bibinfo {author} {\bibfnamefont {M.}~\bibnamefont
  {Zaldarriaga}},\ }\href {\doibase 10.1103/PhysRevLett.88.091303} {\bibfield
  {journal} {\bibinfo  {journal} {Phys. Rev. Lett.}\ }\textbf {\bibinfo
  {volume} {88}},\ \bibinfo {pages} {091303} (\bibinfo {year} {2002})},\
  \Eprint {http://arxiv.org/abs/hep-ph/0108217} {arXiv:hep-ph/0108217}
  \BibitemShut {NoStop}%
\bibitem [{\citenamefont {Olive}\ and\ \citenamefont
  {Pospelov}(2008)}]{Olive:2007aj}%
  \BibitemOpen
  \bibfield  {author} {\bibinfo {author} {\bibfnamefont {K.~A.}\ \bibnamefont
  {Olive}}\ and\ \bibinfo {author} {\bibfnamefont {M.}~\bibnamefont
  {Pospelov}},\ }\href {\doibase 10.1103/PhysRevD.77.043524} {\bibfield
  {journal} {\bibinfo  {journal} {Phys. Rev. D}\ }\textbf {\bibinfo {volume}
  {77}},\ \bibinfo {pages} {043524} (\bibinfo {year} {2008})},\ \Eprint
  {http://arxiv.org/abs/0709.3825} {arXiv:0709.3825 [hep-ph]} \BibitemShut
  {NoStop}%
\bibitem [{\citenamefont {Damour}(2012)}]{Damour:2012rc}%
  \BibitemOpen
  \bibfield  {author} {\bibinfo {author} {\bibfnamefont {T.}~\bibnamefont
  {Damour}},\ }\href {\doibase 10.1088/0264-9381/29/18/184001} {\bibfield
  {journal} {\bibinfo  {journal} {Class. Quant. Grav.}\ }\textbf {\bibinfo
  {volume} {29}},\ \bibinfo {pages} {184001} (\bibinfo {year} {2012})},\
  \Eprint {http://arxiv.org/abs/1202.6311} {arXiv:1202.6311 [gr-qc]}
  \BibitemShut {NoStop}%
\bibitem [{\citenamefont {Armendariz-Picon}(2002)}]{ArmendarizPicon:2002qb}%
  \BibitemOpen
  \bibfield  {author} {\bibinfo {author} {\bibfnamefont {C.}~\bibnamefont
  {Armendariz-Picon}},\ }\href {\doibase 10.1103/PhysRevD.66.064008} {\bibfield
   {journal} {\bibinfo  {journal} {Phys. Rev. D}\ }\textbf {\bibinfo {volume}
  {66}},\ \bibinfo {pages} {064008} (\bibinfo {year} {2002})},\ \Eprint
  {http://arxiv.org/abs/astro-ph/0205187} {arXiv:astro-ph/0205187} \BibitemShut
  {NoStop}%
\bibitem [{\citenamefont {Adelberger}\ \emph {et~al.}(2003)\citenamefont
  {Adelberger}, \citenamefont {Heckel},\ and\ \citenamefont
  {Nelson}}]{Adelberger:2003zx}%
  \BibitemOpen
  \bibfield  {author} {\bibinfo {author} {\bibfnamefont {E.~G.}\ \bibnamefont
  {Adelberger}}, \bibinfo {author} {\bibfnamefont {B.~R.}\ \bibnamefont
  {Heckel}}, \ and\ \bibinfo {author} {\bibfnamefont {A.~E.}\ \bibnamefont
  {Nelson}},\ }\href {\doibase 10.1146/annurev.nucl.53.041002.110503}
  {\bibfield  {journal} {\bibinfo  {journal} {Ann. Rev. Nucl. Part. Sci.}\
  }\textbf {\bibinfo {volume} {53}},\ \bibinfo {pages} {77} (\bibinfo {year}
  {2003})},\ \Eprint {http://arxiv.org/abs/hep-ph/0307284}
  {arXiv:hep-ph/0307284} \BibitemShut {NoStop}%
\bibitem [{\citenamefont {Adelberger}\ \emph {et~al.}(2007)\citenamefont
  {Adelberger}, \citenamefont {Heckel}, \citenamefont {Hoedl}, \citenamefont
  {Hoyle}, \citenamefont {Kapner},\ and\ \citenamefont
  {Upadhye}}]{Adelberger:2006dh}%
  \BibitemOpen
  \bibfield  {author} {\bibinfo {author} {\bibfnamefont {E.~G.}\ \bibnamefont
  {Adelberger}}, \bibinfo {author} {\bibfnamefont {B.~R.}\ \bibnamefont
  {Heckel}}, \bibinfo {author} {\bibfnamefont {S.~A.}\ \bibnamefont {Hoedl}},
  \bibinfo {author} {\bibfnamefont {C.~D.}\ \bibnamefont {Hoyle}}, \bibinfo
  {author} {\bibfnamefont {D.~J.}\ \bibnamefont {Kapner}}, \ and\ \bibinfo
  {author} {\bibfnamefont {A.}~\bibnamefont {Upadhye}},\ }\href {\doibase
  10.1103/PhysRevLett.98.131104} {\bibfield  {journal} {\bibinfo  {journal}
  {Phys. Rev. Lett.}\ }\textbf {\bibinfo {volume} {98}},\ \bibinfo {pages}
  {131104} (\bibinfo {year} {2007})},\ \Eprint
  {http://arxiv.org/abs/hep-ph/0611223} {arXiv:hep-ph/0611223} \BibitemShut
  {NoStop}%
\bibitem [{\citenamefont {Adelberger}\ \emph {et~al.}(2009)\citenamefont
  {Adelberger}, \citenamefont {Gundlach}, \citenamefont {Heckel}, \citenamefont
  {Hoedl},\ and\ \citenamefont {Schlamminger}}]{Adelberger:2009zz}%
  \BibitemOpen
  \bibfield  {author} {\bibinfo {author} {\bibfnamefont {E.~G.}\ \bibnamefont
  {Adelberger}}, \bibinfo {author} {\bibfnamefont {J.~H.}\ \bibnamefont
  {Gundlach}}, \bibinfo {author} {\bibfnamefont {B.~R.}\ \bibnamefont
  {Heckel}}, \bibinfo {author} {\bibfnamefont {S.}~\bibnamefont {Hoedl}}, \
  and\ \bibinfo {author} {\bibfnamefont {S.}~\bibnamefont {Schlamminger}},\
  }\href {\doibase 10.1016/j.ppnp.2008.08.002} {\bibfield  {journal} {\bibinfo
  {journal} {Prog. Part. Nucl. Phys.}\ }\textbf {\bibinfo {volume} {62}},\
  \bibinfo {pages} {102} (\bibinfo {year} {2009})}\BibitemShut {NoStop}%
\bibitem [{\citenamefont {Kapner}\ \emph {et~al.}(2007)\citenamefont {Kapner},
  \citenamefont {Cook}, \citenamefont {Adelberger}, \citenamefont {Gundlach},
  \citenamefont {Heckel}, \citenamefont {Hoyle},\ and\ \citenamefont
  {Swanson}}]{Kapner:2006si}%
  \BibitemOpen
  \bibfield  {author} {\bibinfo {author} {\bibfnamefont {D.~J.}\ \bibnamefont
  {Kapner}}, \bibinfo {author} {\bibfnamefont {T.~S.}\ \bibnamefont {Cook}},
  \bibinfo {author} {\bibfnamefont {E.~G.}\ \bibnamefont {Adelberger}},
  \bibinfo {author} {\bibfnamefont {J.~H.}\ \bibnamefont {Gundlach}}, \bibinfo
  {author} {\bibfnamefont {B.~R.}\ \bibnamefont {Heckel}}, \bibinfo {author}
  {\bibfnamefont {C.~D.}\ \bibnamefont {Hoyle}}, \ and\ \bibinfo {author}
  {\bibfnamefont {H.~E.}\ \bibnamefont {Swanson}},\ }\href {\doibase
  10.1103/PhysRevLett.98.021101} {\bibfield  {journal} {\bibinfo  {journal}
  {Phys. Rev. Lett.}\ }\textbf {\bibinfo {volume} {98}},\ \bibinfo {pages}
  {021101} (\bibinfo {year} {2007})},\ \Eprint
  {http://arxiv.org/abs/hep-ph/0611184} {arXiv:hep-ph/0611184} \BibitemShut
  {NoStop}%
\bibitem [{\citenamefont {Will}(2006)}]{Will:2005va}%
  \BibitemOpen
  \bibfield  {author} {\bibinfo {author} {\bibfnamefont {C.~M.}\ \bibnamefont
  {Will}},\ }\href {\doibase 10.12942/lrr-2006-3} {\bibfield  {journal}
  {\bibinfo  {journal} {Living Rev. Rel.}\ }\textbf {\bibinfo {volume} {9}},\
  \bibinfo {pages} {3} (\bibinfo {year} {2006})},\ \Eprint
  {http://arxiv.org/abs/gr-qc/0510072} {arXiv:gr-qc/0510072} \BibitemShut
  {NoStop}%
\bibitem [{\citenamefont {Rosenband}\ \emph {et~al.}(2008)\citenamefont
  {Rosenband}, \citenamefont {Hume}, \citenamefont {Schmidt}, \citenamefont
  {Chou}, \citenamefont {Brusch}, \citenamefont {Lorini}, \citenamefont
  {Oskay}, \citenamefont {Drullinger}, \citenamefont {Fortier}, \citenamefont
  {Stalnaker}, \citenamefont {Diddams}, \citenamefont {Swann}, \citenamefont
  {Newbury}, \citenamefont {Itano}, \citenamefont {Wineland},\ and\
  \citenamefont {Bergquist}}]{Rosenband1808}%
  \BibitemOpen
  \bibfield  {author} {\bibinfo {author} {\bibfnamefont {T.}~\bibnamefont
  {Rosenband}}, \bibinfo {author} {\bibfnamefont {D.~B.}\ \bibnamefont {Hume}},
  \bibinfo {author} {\bibfnamefont {P.~O.}\ \bibnamefont {Schmidt}}, \bibinfo
  {author} {\bibfnamefont {C.~W.}\ \bibnamefont {Chou}}, \bibinfo {author}
  {\bibfnamefont {A.}~\bibnamefont {Brusch}}, \bibinfo {author} {\bibfnamefont
  {L.}~\bibnamefont {Lorini}}, \bibinfo {author} {\bibfnamefont {W.~H.}\
  \bibnamefont {Oskay}}, \bibinfo {author} {\bibfnamefont {R.~E.}\ \bibnamefont
  {Drullinger}}, \bibinfo {author} {\bibfnamefont {T.~M.}\ \bibnamefont
  {Fortier}}, \bibinfo {author} {\bibfnamefont {J.~E.}\ \bibnamefont
  {Stalnaker}}, \bibinfo {author} {\bibfnamefont {S.~A.}\ \bibnamefont
  {Diddams}}, \bibinfo {author} {\bibfnamefont {W.~C.}\ \bibnamefont {Swann}},
  \bibinfo {author} {\bibfnamefont {N.~R.}\ \bibnamefont {Newbury}}, \bibinfo
  {author} {\bibfnamefont {W.~M.}\ \bibnamefont {Itano}}, \bibinfo {author}
  {\bibfnamefont {D.~J.}\ \bibnamefont {Wineland}}, \ and\ \bibinfo {author}
  {\bibfnamefont {J.~C.}\ \bibnamefont {Bergquist}},\ }\href {\doibase
  10.1126/science.1154622} {\bibfield  {journal} {\bibinfo  {journal}
  {Science}\ }\textbf {\bibinfo {volume} {319}},\ \bibinfo {pages} {1808}
  (\bibinfo {year} {2008})},\ \Eprint
  {http://arxiv.org/abs/https://science.sciencemag.org/content/319/5871/1808.full.pdf}
  {https://science.sciencemag.org/content/319/5871/1808.full.pdf} \BibitemShut
  {NoStop}%
\bibitem [{\citenamefont {Williams}\ \emph {et~al.}(2012)\citenamefont
  {Williams}, \citenamefont {Turyshev},\ and\ \citenamefont
  {Boggs}}]{Williams:2012nc}%
  \BibitemOpen
  \bibfield  {author} {\bibinfo {author} {\bibfnamefont {J.~G.}\ \bibnamefont
  {Williams}}, \bibinfo {author} {\bibfnamefont {S.~G.}\ \bibnamefont
  {Turyshev}}, \ and\ \bibinfo {author} {\bibfnamefont {D.}~\bibnamefont
  {Boggs}},\ }\href {\doibase 10.1088/0264-9381/29/18/184004} {\bibfield
  {journal} {\bibinfo  {journal} {Class. Quant. Grav.}\ }\textbf {\bibinfo
  {volume} {29}},\ \bibinfo {pages} {184004} (\bibinfo {year} {2012})},\
  \Eprint {http://arxiv.org/abs/1203.2150} {arXiv:1203.2150 [gr-qc]}
  \BibitemShut {NoStop}%
\bibitem [{\citenamefont {Tseytlin}\ and\ \citenamefont
  {Vafa}(1992)}]{Tseytlin:1991xk}%
  \BibitemOpen
  \bibfield  {author} {\bibinfo {author} {\bibfnamefont {A.~A.}\ \bibnamefont
  {Tseytlin}}\ and\ \bibinfo {author} {\bibfnamefont {C.}~\bibnamefont
  {Vafa}},\ }\href {\doibase 10.1016/0550-3213(92)90327-8} {\bibfield
  {journal} {\bibinfo  {journal} {Nucl. Phys. B}\ }\textbf {\bibinfo {volume}
  {372}},\ \bibinfo {pages} {443} (\bibinfo {year} {1992})},\ \Eprint
  {http://arxiv.org/abs/hep-th/9109048} {arXiv:hep-th/9109048} \BibitemShut
  {NoStop}%
\bibitem [{\citenamefont {Damour}\ and\ \citenamefont
  {Vilenkin}(1996)}]{Damour:1995pd}%
  \BibitemOpen
  \bibfield  {author} {\bibinfo {author} {\bibfnamefont {T.}~\bibnamefont
  {Damour}}\ and\ \bibinfo {author} {\bibfnamefont {A.}~\bibnamefont
  {Vilenkin}},\ }\href {\doibase 10.1103/PhysRevD.53.2981} {\bibfield
  {journal} {\bibinfo  {journal} {Phys. Rev. D}\ }\textbf {\bibinfo {volume}
  {53}},\ \bibinfo {pages} {2981} (\bibinfo {year} {1996})},\ \Eprint
  {http://arxiv.org/abs/hep-th/9503149} {arXiv:hep-th/9503149} \BibitemShut
  {NoStop}%
\bibitem [{\citenamefont {Damour}\ \emph {et~al.}(2002)\citenamefont {Damour},
  \citenamefont {Piazza},\ and\ \citenamefont {Veneziano}}]{Damour:2002mi}%
  \BibitemOpen
  \bibfield  {author} {\bibinfo {author} {\bibfnamefont {T.}~\bibnamefont
  {Damour}}, \bibinfo {author} {\bibfnamefont {F.}~\bibnamefont {Piazza}}, \
  and\ \bibinfo {author} {\bibfnamefont {G.}~\bibnamefont {Veneziano}},\ }\href
  {\doibase 10.1103/PhysRevLett.89.081601} {\bibfield  {journal} {\bibinfo
  {journal} {Phys. Rev. Lett.}\ }\textbf {\bibinfo {volume} {89}},\ \bibinfo
  {pages} {081601} (\bibinfo {year} {2002})},\ \Eprint
  {http://arxiv.org/abs/gr-qc/0204094} {arXiv:gr-qc/0204094} \BibitemShut
  {NoStop}%
\bibitem [{\citenamefont {Jarv}\ \emph {et~al.}(2008)\citenamefont {Jarv},
  \citenamefont {Kuusk},\ and\ \citenamefont {Saal}}]{Jarv:2008eb}%
  \BibitemOpen
  \bibfield  {author} {\bibinfo {author} {\bibfnamefont {L.}~\bibnamefont
  {Jarv}}, \bibinfo {author} {\bibfnamefont {P.}~\bibnamefont {Kuusk}}, \ and\
  \bibinfo {author} {\bibfnamefont {M.}~\bibnamefont {Saal}},\ }\href {\doibase
  10.1103/PhysRevD.78.083530} {\bibfield  {journal} {\bibinfo  {journal} {Phys.
  Rev. D}\ }\textbf {\bibinfo {volume} {78}},\ \bibinfo {pages} {083530}
  (\bibinfo {year} {2008})},\ \Eprint {http://arxiv.org/abs/0807.2159}
  {arXiv:0807.2159 [gr-qc]} \BibitemShut {NoStop}%
\bibitem [{\citenamefont {Damour}\ and\ \citenamefont
  {Nordtvedt}(1993)}]{Damour:1992kf}%
  \BibitemOpen
  \bibfield  {author} {\bibinfo {author} {\bibfnamefont {T.}~\bibnamefont
  {Damour}}\ and\ \bibinfo {author} {\bibfnamefont {K.}~\bibnamefont
  {Nordtvedt}},\ }\href {\doibase 10.1103/PhysRevLett.70.2217} {\bibfield
  {journal} {\bibinfo  {journal} {Phys. Rev. Lett.}\ }\textbf {\bibinfo
  {volume} {70}},\ \bibinfo {pages} {2217} (\bibinfo {year}
  {1993})}\BibitemShut {NoStop}%
\bibitem [{\citenamefont {Khoury}(2010)}]{Khoury:2010xi}%
  \BibitemOpen
  \bibfield  {author} {\bibinfo {author} {\bibfnamefont {J.}~\bibnamefont
  {Khoury}},\ }\href@noop {} {\  (\bibinfo {year} {2010})},\ \Eprint
  {http://arxiv.org/abs/1011.5909} {arXiv:1011.5909 [astro-ph.CO]} \BibitemShut
  {NoStop}%
\bibitem [{\citenamefont {Khoury}\ and\ \citenamefont
  {Weltman}(2004{\natexlab{a}})}]{Khoury:2003aq}%
  \BibitemOpen
  \bibfield  {author} {\bibinfo {author} {\bibfnamefont {J.}~\bibnamefont
  {Khoury}}\ and\ \bibinfo {author} {\bibfnamefont {A.}~\bibnamefont
  {Weltman}},\ }\href {\doibase 10.1103/PhysRevLett.93.171104} {\bibfield
  {journal} {\bibinfo  {journal} {Phys. Rev. Lett.}\ }\textbf {\bibinfo
  {volume} {93}},\ \bibinfo {pages} {171104} (\bibinfo {year}
  {2004}{\natexlab{a}})},\ \Eprint {http://arxiv.org/abs/astro-ph/0309300}
  {arXiv:astro-ph/0309300} \BibitemShut {NoStop}%
\bibitem [{\citenamefont {Khoury}\ and\ \citenamefont
  {Weltman}(2004{\natexlab{b}})}]{Khoury:2003rn}%
  \BibitemOpen
  \bibfield  {author} {\bibinfo {author} {\bibfnamefont {J.}~\bibnamefont
  {Khoury}}\ and\ \bibinfo {author} {\bibfnamefont {A.}~\bibnamefont
  {Weltman}},\ }\href {\doibase 10.1103/PhysRevD.69.044026} {\bibfield
  {journal} {\bibinfo  {journal} {Phys. Rev. D}\ }\textbf {\bibinfo {volume}
  {69}},\ \bibinfo {pages} {044026} (\bibinfo {year} {2004}{\natexlab{b}})},\
  \Eprint {http://arxiv.org/abs/astro-ph/0309411} {arXiv:astro-ph/0309411}
  \BibitemShut {NoStop}%
\bibitem [{\citenamefont {Hees}\ and\ \citenamefont
  {Fuzfa}(2012)}]{Hees:2011mu}%
  \BibitemOpen
  \bibfield  {author} {\bibinfo {author} {\bibfnamefont {A.}~\bibnamefont
  {Hees}}\ and\ \bibinfo {author} {\bibfnamefont {A.}~\bibnamefont {Fuzfa}},\
  }\href {\doibase 10.1103/PhysRevD.85.103005} {\bibfield  {journal} {\bibinfo
  {journal} {Phys. Rev. D}\ }\textbf {\bibinfo {volume} {85}},\ \bibinfo
  {pages} {103005} (\bibinfo {year} {2012})},\ \Eprint
  {http://arxiv.org/abs/1111.4784} {arXiv:1111.4784 [gr-qc]} \BibitemShut
  {NoStop}%
\bibitem [{\citenamefont {Hinterbichler}\ and\ \citenamefont
  {Khoury}(2010)}]{Hinterbichler:2010es}%
  \BibitemOpen
  \bibfield  {author} {\bibinfo {author} {\bibfnamefont {K.}~\bibnamefont
  {Hinterbichler}}\ and\ \bibinfo {author} {\bibfnamefont {J.}~\bibnamefont
  {Khoury}},\ }\href {\doibase 10.1103/PhysRevLett.104.231301} {\bibfield
  {journal} {\bibinfo  {journal} {Phys. Rev. Lett.}\ }\textbf {\bibinfo
  {volume} {104}},\ \bibinfo {pages} {231301} (\bibinfo {year} {2010})},\
  \Eprint {http://arxiv.org/abs/1001.4525} {arXiv:1001.4525 [hep-th]}
  \BibitemShut {NoStop}%
\bibitem [{\citenamefont {Hinterbichler}\ \emph {et~al.}(2011)\citenamefont
  {Hinterbichler}, \citenamefont {Khoury}, \citenamefont {Levy},\ and\
  \citenamefont {Matas}}]{Hinterbichler:2011ca}%
  \BibitemOpen
  \bibfield  {author} {\bibinfo {author} {\bibfnamefont {K.}~\bibnamefont
  {Hinterbichler}}, \bibinfo {author} {\bibfnamefont {J.}~\bibnamefont
  {Khoury}}, \bibinfo {author} {\bibfnamefont {A.}~\bibnamefont {Levy}}, \ and\
  \bibinfo {author} {\bibfnamefont {A.}~\bibnamefont {Matas}},\ }\href
  {\doibase 10.1103/PhysRevD.84.103521} {\bibfield  {journal} {\bibinfo
  {journal} {Phys. Rev. D}\ }\textbf {\bibinfo {volume} {84}},\ \bibinfo
  {pages} {103521} (\bibinfo {year} {2011})},\ \Eprint
  {http://arxiv.org/abs/1107.2112} {arXiv:1107.2112 [astro-ph.CO]} \BibitemShut
  {NoStop}%
\bibitem [{\citenamefont {Webb}\ \emph {et~al.}(2001)\citenamefont {Webb},
  \citenamefont {Murphy}, \citenamefont {Flambaum}, \citenamefont {Dzuba},
  \citenamefont {Barrow}, \citenamefont {Churchill}, \citenamefont
  {Prochaska},\ and\ \citenamefont {Wolfe}}]{Webb:2000mn}%
  \BibitemOpen
  \bibfield  {author} {\bibinfo {author} {\bibfnamefont {J.~K.}\ \bibnamefont
  {Webb}}, \bibinfo {author} {\bibfnamefont {M.~T.}\ \bibnamefont {Murphy}},
  \bibinfo {author} {\bibfnamefont {V.~V.}\ \bibnamefont {Flambaum}}, \bibinfo
  {author} {\bibfnamefont {V.~A.}\ \bibnamefont {Dzuba}}, \bibinfo {author}
  {\bibfnamefont {J.~D.}\ \bibnamefont {Barrow}}, \bibinfo {author}
  {\bibfnamefont {C.~W.}\ \bibnamefont {Churchill}}, \bibinfo {author}
  {\bibfnamefont {J.~X.}\ \bibnamefont {Prochaska}}, \ and\ \bibinfo {author}
  {\bibfnamefont {A.~M.}\ \bibnamefont {Wolfe}},\ }\href {\doibase
  10.1103/PhysRevLett.87.091301} {\bibfield  {journal} {\bibinfo  {journal}
  {Phys. Rev. Lett.}\ }\textbf {\bibinfo {volume} {87}},\ \bibinfo {pages}
  {091301} (\bibinfo {year} {2001})},\ \Eprint
  {http://arxiv.org/abs/astro-ph/0012539} {arXiv:astro-ph/0012539} \BibitemShut
  {NoStop}%
\bibitem [{\citenamefont {King}\ \emph {et~al.}(2012)\citenamefont {King},
  \citenamefont {Webb}, \citenamefont {Murphy}, \citenamefont {Flambaum},
  \citenamefont {Carswell}, \citenamefont {Bainbridge}, \citenamefont
  {Wilczynska},\ and\ \citenamefont {Koch}}]{King:2012id}%
  \BibitemOpen
  \bibfield  {author} {\bibinfo {author} {\bibfnamefont {J.~A.}\ \bibnamefont
  {King}}, \bibinfo {author} {\bibfnamefont {J.~K.}\ \bibnamefont {Webb}},
  \bibinfo {author} {\bibfnamefont {M.~T.}\ \bibnamefont {Murphy}}, \bibinfo
  {author} {\bibfnamefont {V.~V.}\ \bibnamefont {Flambaum}}, \bibinfo {author}
  {\bibfnamefont {R.~F.}\ \bibnamefont {Carswell}}, \bibinfo {author}
  {\bibfnamefont {M.~B.}\ \bibnamefont {Bainbridge}}, \bibinfo {author}
  {\bibfnamefont {M.~R.}\ \bibnamefont {Wilczynska}}, \ and\ \bibinfo {author}
  {\bibfnamefont {F.~E.}\ \bibnamefont {Koch}},\ }\href {\doibase
  10.1111/j.1365-2966.2012.20852.x} {\bibfield  {journal} {\bibinfo  {journal}
  {Mon. Not. Roy. Astron. Soc.}\ }\textbf {\bibinfo {volume} {422}},\ \bibinfo
  {pages} {3370} (\bibinfo {year} {2012})},\ \Eprint
  {http://arxiv.org/abs/1202.4758} {arXiv:1202.4758 [astro-ph.CO]} \BibitemShut
  {NoStop}%
\bibitem [{\citenamefont {Webb}\ \emph {et~al.}(2011)\citenamefont {Webb},
  \citenamefont {King}, \citenamefont {Murphy}, \citenamefont {Flambaum},
  \citenamefont {Carswell},\ and\ \citenamefont {Bainbridge}}]{Webb:2010hc}%
  \BibitemOpen
  \bibfield  {author} {\bibinfo {author} {\bibfnamefont {J.~K.}\ \bibnamefont
  {Webb}}, \bibinfo {author} {\bibfnamefont {J.~A.}\ \bibnamefont {King}},
  \bibinfo {author} {\bibfnamefont {M.~T.}\ \bibnamefont {Murphy}}, \bibinfo
  {author} {\bibfnamefont {V.~V.}\ \bibnamefont {Flambaum}}, \bibinfo {author}
  {\bibfnamefont {R.~F.}\ \bibnamefont {Carswell}}, \ and\ \bibinfo {author}
  {\bibfnamefont {M.~B.}\ \bibnamefont {Bainbridge}},\ }\href {\doibase
  10.1103/PhysRevLett.107.191101} {\bibfield  {journal} {\bibinfo  {journal}
  {Phys. Rev. Lett.}\ }\textbf {\bibinfo {volume} {107}},\ \bibinfo {pages}
  {191101} (\bibinfo {year} {2011})},\ \Eprint {http://arxiv.org/abs/1008.3907}
  {arXiv:1008.3907 [astro-ph.CO]} \BibitemShut {NoStop}%
\bibitem [{\citenamefont {Holanda}\ \emph {et~al.}(2016)\citenamefont
  {Holanda}, \citenamefont {Landau}, \citenamefont {Alcaniz}, \citenamefont
  {Sanchez~G.},\ and\ \citenamefont {Busti}}]{Holanda:2015oda}%
  \BibitemOpen
  \bibfield  {author} {\bibinfo {author} {\bibfnamefont {R.~F.~L.}\
  \bibnamefont {Holanda}}, \bibinfo {author} {\bibfnamefont {S.~J.}\
  \bibnamefont {Landau}}, \bibinfo {author} {\bibfnamefont {J.~S.}\
  \bibnamefont {Alcaniz}}, \bibinfo {author} {\bibfnamefont {I.~E.}\
  \bibnamefont {Sanchez~G.}}, \ and\ \bibinfo {author} {\bibfnamefont {V.~C.}\
  \bibnamefont {Busti}},\ }\href {\doibase 10.1088/1475-7516/2016/05/047}
  {\bibfield  {journal} {\bibinfo  {journal} {JCAP}\ }\textbf {\bibinfo
  {volume} {05}},\ \bibinfo {pages} {047} (\bibinfo {year} {2016})},\ \Eprint
  {http://arxiv.org/abs/1510.07240} {arXiv:1510.07240 [astro-ph.CO]}
  \BibitemShut {NoStop}%
\bibitem [{\citenamefont {Khatri}\ and\ \citenamefont
  {Wandelt}(2007)}]{Khatri:2007yv}%
  \BibitemOpen
  \bibfield  {author} {\bibinfo {author} {\bibfnamefont {R.}~\bibnamefont
  {Khatri}}\ and\ \bibinfo {author} {\bibfnamefont {B.~D.}\ \bibnamefont
  {Wandelt}},\ }\href {\doibase 10.1103/PhysRevLett.98.111301} {\bibfield
  {journal} {\bibinfo  {journal} {Phys. Rev. Lett.}\ }\textbf {\bibinfo
  {volume} {98}},\ \bibinfo {pages} {111301} (\bibinfo {year} {2007})},\
  \Eprint {http://arxiv.org/abs/astro-ph/0701752} {arXiv:astro-ph/0701752}
  \BibitemShut {NoStop}%
\bibitem [{\citenamefont {Riess}\ \emph {et~al.}(2016)\citenamefont {Riess}
  \emph {et~al.}}]{Riess:2016jrr}%
  \BibitemOpen
  \bibfield  {author} {\bibinfo {author} {\bibfnamefont {A.~G.}\ \bibnamefont
  {Riess}} \emph {et~al.},\ }\href {\doibase 10.3847/0004-637X/826/1/56}
  {\bibfield  {journal} {\bibinfo  {journal} {Astrophys. J.}\ }\textbf
  {\bibinfo {volume} {826}},\ \bibinfo {pages} {56} (\bibinfo {year} {2016})},\
  \Eprint {http://arxiv.org/abs/1604.01424} {arXiv:1604.01424 [astro-ph.CO]}
  \BibitemShut {NoStop}%
\bibitem [{\citenamefont {{Cao}}\ and\ \citenamefont
  {{Liang}}(2011)}]{2011RAA....11.1199C}%
  \BibitemOpen
  \bibfield  {author} {\bibinfo {author} {\bibfnamefont {S.}~\bibnamefont
  {{Cao}}}\ and\ \bibinfo {author} {\bibfnamefont {N.}~\bibnamefont
  {{Liang}}},\ }\href {\doibase 10.1088/1674-4527/11/10/008} {\bibfield
  {journal} {\bibinfo  {journal} {Research in Astronomy and Astrophysics}\
  }\textbf {\bibinfo {volume} {11}},\ \bibinfo {pages} {1199} (\bibinfo {year}
  {2011})},\ \Eprint {http://arxiv.org/abs/1104.4942} {arXiv:1104.4942
  [astro-ph.CO]} \BibitemShut {NoStop}%
\bibitem [{\citenamefont {Hees}\ \emph {et~al.}(2014)\citenamefont {Hees},
  \citenamefont {Minazzoli},\ and\ \citenamefont {Larena}}]{Hees:2014lfa}%
  \BibitemOpen
  \bibfield  {author} {\bibinfo {author} {\bibfnamefont {A.}~\bibnamefont
  {Hees}}, \bibinfo {author} {\bibfnamefont {O.}~\bibnamefont {Minazzoli}}, \
  and\ \bibinfo {author} {\bibfnamefont {J.}~\bibnamefont {Larena}},\ }\href
  {\doibase 10.1103/PhysRevD.90.124064} {\bibfield  {journal} {\bibinfo
  {journal} {Phys. Rev. D}\ }\textbf {\bibinfo {volume} {90}},\ \bibinfo
  {pages} {124064} (\bibinfo {year} {2014})},\ \Eprint
  {http://arxiv.org/abs/1406.6187} {arXiv:1406.6187 [astro-ph.CO]} \BibitemShut
  {NoStop}%
\bibitem [{\citenamefont {Abbott}\ \emph
  {et~al.}(2017{\natexlab{a}})\citenamefont {Abbott} \emph
  {et~al.}}]{LIGOScientific:2017vwq}%
  \BibitemOpen
  \bibfield  {author} {\bibinfo {author} {\bibfnamefont {B.~P.}\ \bibnamefont
  {Abbott}} \emph {et~al.} (\bibinfo {collaboration} {LIGO Scientific,
  Virgo}),\ }\href {\doibase 10.1103/PhysRevLett.119.161101} {\bibfield
  {journal} {\bibinfo  {journal} {Phys. Rev. Lett.}\ }\textbf {\bibinfo
  {volume} {119}},\ \bibinfo {pages} {161101} (\bibinfo {year}
  {2017}{\natexlab{a}})},\ \Eprint {http://arxiv.org/abs/1710.05832}
  {arXiv:1710.05832 [gr-qc]} \BibitemShut {NoStop}%
\bibitem [{\citenamefont {Abbott}\ \emph
  {et~al.}(2017{\natexlab{b}})\citenamefont {Abbott} \emph
  {et~al.}}]{LIGOScientific:2017ync}%
  \BibitemOpen
  \bibfield  {author} {\bibinfo {author} {\bibfnamefont {B.~P.}\ \bibnamefont
  {Abbott}} \emph {et~al.} (\bibinfo {collaboration} {LIGO Scientific, Virgo,
  Fermi GBM, INTEGRAL, IceCube, AstroSat Cadmium Zinc Telluride Imager Team,
  IPN, Insight-Hxmt, ANTARES, Swift, AGILE Team, 1M2H Team, Dark Energy Camera
  GW-EM, DES, DLT40, GRAWITA, Fermi-LAT, ATCA, ASKAP, Las Cumbres Observatory
  Group, OzGrav, DWF (Deeper Wider Faster Program), AST3, CAASTRO, VINROUGE,
  MASTER, J-GEM, GROWTH, JAGWAR, CaltechNRAO, TTU-NRAO, NuSTAR, Pan-STARRS,
  MAXI Team, TZAC Consortium, KU, Nordic Optical Telescope, ePESSTO, GROND,
  Texas Tech University, SALT Group, TOROS, BOOTES, MWA, CALET, IKI-GW
  Follow-up, H.E.S.S., LOFAR, LWA, HAWC, Pierre Auger, ALMA, Euro VLBI Team, Pi
  of Sky, Chandra Team at McGill University, DFN, ATLAS Telescopes, High Time
  Resolution Universe Survey, RIMAS, RATIR, SKA South Africa/MeerKAT}),\ }\href
  {\doibase 10.3847/2041-8213/aa91c9} {\bibfield  {journal} {\bibinfo
  {journal} {Astrophys. J. Lett.}\ }\textbf {\bibinfo {volume} {848}},\
  \bibinfo {pages} {L12} (\bibinfo {year} {2017}{\natexlab{b}})},\ \Eprint
  {http://arxiv.org/abs/1710.05833} {arXiv:1710.05833 [astro-ph.HE]}
  \BibitemShut {NoStop}%
\bibitem [{\citenamefont {Fanizza}\ \emph {et~al.}(2020)\citenamefont
  {Fanizza}, \citenamefont {Franchini}, \citenamefont {Gasperini},\ and\
  \citenamefont {Tedesco}}]{Fanizza:2020hat}%
  \BibitemOpen
  \bibfield  {author} {\bibinfo {author} {\bibfnamefont {G.}~\bibnamefont
  {Fanizza}}, \bibinfo {author} {\bibfnamefont {G.}~\bibnamefont {Franchini}},
  \bibinfo {author} {\bibfnamefont {M.}~\bibnamefont {Gasperini}}, \ and\
  \bibinfo {author} {\bibfnamefont {L.}~\bibnamefont {Tedesco}},\ }\href
  {\doibase 10.1007/s10714-020-02760-5} {\bibfield  {journal} {\bibinfo
  {journal} {Gen. Rel. Grav.}\ }\textbf {\bibinfo {volume} {52}},\ \bibinfo
  {pages} {111} (\bibinfo {year} {2020})},\ \Eprint
  {http://arxiv.org/abs/2010.06569} {arXiv:2010.06569 [gr-qc]} \BibitemShut
  {NoStop}%
\bibitem [{\citenamefont {Finke}\ \emph {et~al.}(2021)\citenamefont {Finke},
  \citenamefont {Foffa}, \citenamefont {Iacovelli}, \citenamefont {Maggiore},\
  and\ \citenamefont {Mancarella}}]{Finke:2021aom}%
  \BibitemOpen
  \bibfield  {author} {\bibinfo {author} {\bibfnamefont {A.}~\bibnamefont
  {Finke}}, \bibinfo {author} {\bibfnamefont {S.}~\bibnamefont {Foffa}},
  \bibinfo {author} {\bibfnamefont {F.}~\bibnamefont {Iacovelli}}, \bibinfo
  {author} {\bibfnamefont {M.}~\bibnamefont {Maggiore}}, \ and\ \bibinfo
  {author} {\bibfnamefont {M.}~\bibnamefont {Mancarella}},\ }\href@noop {} {\
  (\bibinfo {year} {2021})},\ \Eprint {http://arxiv.org/abs/2101.12660}
  {arXiv:2101.12660 [astro-ph.CO]} \BibitemShut {NoStop}%
\bibitem [{\citenamefont {Mukherjee}\ \emph {et~al.}(2020)\citenamefont
  {Mukherjee}, \citenamefont {Wandelt},\ and\ \citenamefont
  {Silk}}]{Mukherjee:2020mha}%
  \BibitemOpen
  \bibfield  {author} {\bibinfo {author} {\bibfnamefont {S.}~\bibnamefont
  {Mukherjee}}, \bibinfo {author} {\bibfnamefont {B.~D.}\ \bibnamefont
  {Wandelt}}, \ and\ \bibinfo {author} {\bibfnamefont {J.}~\bibnamefont
  {Silk}},\ }\href {\doibase 10.1093/mnras/stab001} {\  (\bibinfo {year}
  {2020}),\ 10.1093/mnras/stab001},\ \Eprint {http://arxiv.org/abs/2012.15316}
  {arXiv:2012.15316 [astro-ph.CO]} \BibitemShut {NoStop}%
\bibitem [{\citenamefont {Minazzoli}\ and\ \citenamefont
  {Hees}(2014)}]{Minazzoli:2014xua}%
  \BibitemOpen
  \bibfield  {author} {\bibinfo {author} {\bibfnamefont {O.}~\bibnamefont
  {Minazzoli}}\ and\ \bibinfo {author} {\bibfnamefont {A.}~\bibnamefont
  {Hees}},\ }\href {\doibase 10.1103/PhysRevD.90.023017} {\bibfield  {journal}
  {\bibinfo  {journal} {Phys. Rev. D}\ }\textbf {\bibinfo {volume} {90}},\
  \bibinfo {pages} {023017} (\bibinfo {year} {2014})},\ \Eprint
  {http://arxiv.org/abs/1404.4266} {arXiv:1404.4266 [gr-qc]} \BibitemShut
  {NoStop}%
\bibitem [{\citenamefont {Reitze}\ \emph {et~al.}(2019)\citenamefont {Reitze}
  \emph {et~al.}}]{Reitze:2019iox}%
  \BibitemOpen
  \bibfield  {author} {\bibinfo {author} {\bibfnamefont {D.}~\bibnamefont
  {Reitze}} \emph {et~al.},\ }\href@noop {} {\bibfield  {journal} {\bibinfo
  {journal} {Bull. Am. Astron. Soc.}\ }\textbf {\bibinfo {volume} {51}},\
  \bibinfo {pages} {035} (\bibinfo {year} {2019})},\ \Eprint
  {http://arxiv.org/abs/1907.04833} {arXiv:1907.04833 [astro-ph.IM]}
  \BibitemShut {NoStop}%
\bibitem [{\citenamefont {Abbott}\ \emph {et~al.}(2018)\citenamefont {Abbott}
  \emph {et~al.}}]{KAGRA:2013rdx}%
  \BibitemOpen
  \bibfield  {author} {\bibinfo {author} {\bibfnamefont {B.~P.}\ \bibnamefont
  {Abbott}} \emph {et~al.} (\bibinfo {collaboration} {KAGRA, LIGO Scientific,
  Virgo, VIRGO}),\ }\href {\doibase 10.1007/s41114-020-00026-9} {\bibfield
  {journal} {\bibinfo  {journal} {Living Rev. Rel.}\ }\textbf {\bibinfo
  {volume} {21}},\ \bibinfo {pages} {3} (\bibinfo {year} {2018})},\ \Eprint
  {http://arxiv.org/abs/1304.0670} {arXiv:1304.0670 [gr-qc]} \BibitemShut
  {NoStop}%
\bibitem [{\citenamefont {Acernese}\ \emph
  {et~al.}(2015{\natexlab{a}})\citenamefont {Acernese} \emph
  {et~al.}}]{VIRGO:2014yos}%
  \BibitemOpen
  \bibfield  {author} {\bibinfo {author} {\bibfnamefont {F.}~\bibnamefont
  {Acernese}} \emph {et~al.} (\bibinfo {collaboration} {VIRGO}),\ }\href
  {\doibase 10.1088/0264-9381/32/2/024001} {\bibfield  {journal} {\bibinfo
  {journal} {Class. Quant. Grav.}\ }\textbf {\bibinfo {volume} {32}},\ \bibinfo
  {pages} {024001} (\bibinfo {year} {2015}{\natexlab{a}})},\ \Eprint
  {http://arxiv.org/abs/1408.3978} {arXiv:1408.3978 [gr-qc]} \BibitemShut
  {NoStop}%
\bibitem [{\citenamefont {Aso}\ \emph {et~al.}(2013)\citenamefont {Aso},
  \citenamefont {Michimura}, \citenamefont {Somiya}, \citenamefont {Ando},
  \citenamefont {Miyakawa}, \citenamefont {Sekiguchi}, \citenamefont
  {Tatsumi},\ and\ \citenamefont {Yamamoto}}]{Aso:2013eba}%
  \BibitemOpen
  \bibfield  {author} {\bibinfo {author} {\bibfnamefont {Y.}~\bibnamefont
  {Aso}}, \bibinfo {author} {\bibfnamefont {Y.}~\bibnamefont {Michimura}},
  \bibinfo {author} {\bibfnamefont {K.}~\bibnamefont {Somiya}}, \bibinfo
  {author} {\bibfnamefont {M.}~\bibnamefont {Ando}}, \bibinfo {author}
  {\bibfnamefont {O.}~\bibnamefont {Miyakawa}}, \bibinfo {author}
  {\bibfnamefont {T.}~\bibnamefont {Sekiguchi}}, \bibinfo {author}
  {\bibfnamefont {D.}~\bibnamefont {Tatsumi}}, \ and\ \bibinfo {author}
  {\bibfnamefont {H.}~\bibnamefont {Yamamoto}} (\bibinfo {collaboration}
  {KAGRA}),\ }\href {\doibase 10.1103/PhysRevD.88.043007} {\bibfield  {journal}
  {\bibinfo  {journal} {Phys. Rev. D}\ }\textbf {\bibinfo {volume} {88}},\
  \bibinfo {pages} {043007} (\bibinfo {year} {2013})},\ \Eprint
  {http://arxiv.org/abs/1306.6747} {arXiv:1306.6747 [gr-qc]} \BibitemShut
  {NoStop}%
\bibitem [{\citenamefont {Somiya}(2012)}]{Somiya:2011np}%
  \BibitemOpen
  \bibfield  {author} {\bibinfo {author} {\bibfnamefont {K.}~\bibnamefont
  {Somiya}} (\bibinfo {collaboration} {KAGRA}),\ }\href {\doibase
  10.1088/0264-9381/29/12/124007} {\bibfield  {journal} {\bibinfo  {journal}
  {Class. Quant. Grav.}\ }\textbf {\bibinfo {volume} {29}},\ \bibinfo {pages}
  {124007} (\bibinfo {year} {2012})},\ \Eprint {http://arxiv.org/abs/1111.7185}
  {arXiv:1111.7185 [gr-qc]} \BibitemShut {NoStop}%
\bibitem [{\citenamefont {Unnikrishnan}(2013)}]{Unnikrishnan:2013qwa}%
  \BibitemOpen
  \bibfield  {author} {\bibinfo {author} {\bibfnamefont {C.~S.}\ \bibnamefont
  {Unnikrishnan}},\ }\href {\doibase 10.1142/S0218271813410101} {\bibfield
  {journal} {\bibinfo  {journal} {Int. J. Mod. Phys. D}\ }\textbf {\bibinfo
  {volume} {22}},\ \bibinfo {pages} {1341010} (\bibinfo {year} {2013})},\
  \Eprint {http://arxiv.org/abs/1510.06059} {arXiv:1510.06059
  [physics.ins-det]} \BibitemShut {NoStop}%
\bibitem [{\citenamefont {Saleem}\ \emph {et~al.}(2022)\citenamefont {Saleem}
  \emph {et~al.}}]{Saleem:2021iwi}%
  \BibitemOpen
  \bibfield  {author} {\bibinfo {author} {\bibfnamefont {M.}~\bibnamefont
  {Saleem}} \emph {et~al.},\ }\href {\doibase 10.1088/1361-6382/ac3b99}
  {\bibfield  {journal} {\bibinfo  {journal} {Class. Quant. Grav.}\ }\textbf
  {\bibinfo {volume} {39}},\ \bibinfo {pages} {025004} (\bibinfo {year}
  {2022})},\ \Eprint {http://arxiv.org/abs/2105.01716} {arXiv:2105.01716
  [gr-qc]} \BibitemShut {NoStop}%
\bibitem [{\citenamefont {Adhikari}\ \emph
  {et~al.}(2020{\natexlab{a}})\citenamefont {Adhikari} \emph
  {et~al.}}]{LIGO:2020xsf}%
  \BibitemOpen
  \bibfield  {author} {\bibinfo {author} {\bibfnamefont {R.~X.}\ \bibnamefont
  {Adhikari}} \emph {et~al.} (\bibinfo {collaboration} {LIGO}),\ }\href
  {\doibase 10.1088/1361-6382/ab9143} {\bibfield  {journal} {\bibinfo
  {journal} {Class. Quant. Grav.}\ }\textbf {\bibinfo {volume} {37}},\ \bibinfo
  {pages} {165003} (\bibinfo {year} {2020}{\natexlab{a}})},\ \Eprint
  {http://arxiv.org/abs/2001.11173} {arXiv:2001.11173 [astro-ph.IM]}
  \BibitemShut {NoStop}%
\bibitem [{\citenamefont {Evans}\ \emph {et~al.}(2021)\citenamefont {Evans}
  \emph {et~al.}}]{Evans:2021gyd}%
  \BibitemOpen
  \bibfield  {author} {\bibinfo {author} {\bibfnamefont {M.}~\bibnamefont
  {Evans}} \emph {et~al.},\ }\href@noop {} {\  (\bibinfo {year} {2021})},\
  \Eprint {http://arxiv.org/abs/2109.09882} {arXiv:2109.09882 [astro-ph.IM]}
  \BibitemShut {NoStop}%
\bibitem [{\citenamefont {Punturo}\ \emph
  {et~al.}(2010{\natexlab{a}})\citenamefont {Punturo} \emph
  {et~al.}}]{Punturo:2010zz}%
  \BibitemOpen
  \bibfield  {author} {\bibinfo {author} {\bibfnamefont {M.}~\bibnamefont
  {Punturo}} \emph {et~al.},\ }\href {\doibase 10.1088/0264-9381/27/19/194002}
  {\bibfield  {journal} {\bibinfo  {journal} {Class. Quant. Grav.}\ }\textbf
  {\bibinfo {volume} {27}},\ \bibinfo {pages} {194002} (\bibinfo {year}
  {2010}{\natexlab{a}})}\BibitemShut {NoStop}%
\bibitem [{\citenamefont {Punturo}\ \emph
  {et~al.}(2010{\natexlab{b}})\citenamefont {Punturo} \emph
  {et~al.}}]{Punturo:2010zza}%
  \BibitemOpen
  \bibfield  {author} {\bibinfo {author} {\bibfnamefont {M.}~\bibnamefont
  {Punturo}} \emph {et~al.},\ }\href {\doibase 10.1088/0264-9381/27/8/084007}
  {\bibfield  {journal} {\bibinfo  {journal} {Class. Quant. Grav.}\ }\textbf
  {\bibinfo {volume} {27}},\ \bibinfo {pages} {084007} (\bibinfo {year}
  {2010}{\natexlab{b}})}\BibitemShut {NoStop}%
\bibitem [{\citenamefont {Hild}\ \emph {et~al.}(2011)\citenamefont {Hild} \emph
  {et~al.}}]{Hild:2010id}%
  \BibitemOpen
  \bibfield  {author} {\bibinfo {author} {\bibfnamefont {S.}~\bibnamefont
  {Hild}} \emph {et~al.},\ }\href {\doibase 10.1088/0264-9381/28/9/094013}
  {\bibfield  {journal} {\bibinfo  {journal} {Class. Quant. Grav.}\ }\textbf
  {\bibinfo {volume} {28}},\ \bibinfo {pages} {094013} (\bibinfo {year}
  {2011})},\ \Eprint {http://arxiv.org/abs/1012.0908} {arXiv:1012.0908 [gr-qc]}
  \BibitemShut {NoStop}%
\bibitem [{\citenamefont {Holanda}\ \emph {et~al.}(2012)\citenamefont
  {Holanda}, \citenamefont {Gon\c{c}alves},\ and\ \citenamefont
  {Alcaniz}}]{Holanda:2012at}%
  \BibitemOpen
  \bibfield  {author} {\bibinfo {author} {\bibfnamefont {R.~F.~L.}\
  \bibnamefont {Holanda}}, \bibinfo {author} {\bibfnamefont {R.~S.}\
  \bibnamefont {Gon\c{c}alves}}, \ and\ \bibinfo {author} {\bibfnamefont
  {J.~S.}\ \bibnamefont {Alcaniz}},\ }\href {\doibase
  10.1088/1475-7516/2012/06/022} {\bibfield  {journal} {\bibinfo  {journal}
  {JCAP}\ }\textbf {\bibinfo {volume} {06}},\ \bibinfo {pages} {022} (\bibinfo
  {year} {2012})},\ \Eprint {http://arxiv.org/abs/1201.2378} {arXiv:1201.2378
  [astro-ph.CO]} \BibitemShut {NoStop}%
\bibitem [{\citenamefont {Yunes}\ \emph {et~al.}(2010)\citenamefont {Yunes},
  \citenamefont {Pretorius},\ and\ \citenamefont {Spergel}}]{Yunes:2009bv}%
  \BibitemOpen
  \bibfield  {author} {\bibinfo {author} {\bibfnamefont {N.}~\bibnamefont
  {Yunes}}, \bibinfo {author} {\bibfnamefont {F.}~\bibnamefont {Pretorius}}, \
  and\ \bibinfo {author} {\bibfnamefont {D.}~\bibnamefont {Spergel}},\ }\href
  {\doibase 10.1103/PhysRevD.81.064018} {\bibfield  {journal} {\bibinfo
  {journal} {Phys. Rev. D}\ }\textbf {\bibinfo {volume} {81}},\ \bibinfo
  {pages} {064018} (\bibinfo {year} {2010})},\ \Eprint
  {http://arxiv.org/abs/0912.2724} {arXiv:0912.2724 [gr-qc]} \BibitemShut
  {NoStop}%
\bibitem [{\citenamefont {Yunes}\ \emph {et~al.}(2016)\citenamefont {Yunes},
  \citenamefont {Yagi},\ and\ \citenamefont {Pretorius}}]{Yunes:2016jcc}%
  \BibitemOpen
  \bibfield  {author} {\bibinfo {author} {\bibfnamefont {N.}~\bibnamefont
  {Yunes}}, \bibinfo {author} {\bibfnamefont {K.}~\bibnamefont {Yagi}}, \ and\
  \bibinfo {author} {\bibfnamefont {F.}~\bibnamefont {Pretorius}},\ }\href
  {\doibase 10.1103/PhysRevD.94.084002} {\bibfield  {journal} {\bibinfo
  {journal} {Phys. Rev. D}\ }\textbf {\bibinfo {volume} {94}},\ \bibinfo
  {pages} {084002} (\bibinfo {year} {2016})},\ \Eprint
  {http://arxiv.org/abs/1603.08955} {arXiv:1603.08955 [gr-qc]} \BibitemShut
  {NoStop}%
\bibitem [{\citenamefont {Vijaykumar}\ \emph {et~al.}(2021)\citenamefont
  {Vijaykumar}, \citenamefont {Kapadia},\ and\ \citenamefont
  {Ajith}}]{Vijaykumar:2020nzc}%
  \BibitemOpen
  \bibfield  {author} {\bibinfo {author} {\bibfnamefont {A.}~\bibnamefont
  {Vijaykumar}}, \bibinfo {author} {\bibfnamefont {S.~J.}\ \bibnamefont
  {Kapadia}}, \ and\ \bibinfo {author} {\bibfnamefont {P.}~\bibnamefont
  {Ajith}},\ }\href {\doibase 10.1103/PhysRevLett.126.141104} {\bibfield
  {journal} {\bibinfo  {journal} {Phys. Rev. Lett.}\ }\textbf {\bibinfo
  {volume} {126}},\ \bibinfo {pages} {141104} (\bibinfo {year} {2021})},\
  \Eprint {http://arxiv.org/abs/2003.12832} {arXiv:2003.12832 [gr-qc]}
  \BibitemShut {NoStop}%
\bibitem [{\citenamefont {Aasi}\ \emph {et~al.}(2015)\citenamefont {Aasi} \emph
  {et~al.}}]{TheLIGOScientific:2014jea}%
  \BibitemOpen
  \bibfield  {author} {\bibinfo {author} {\bibfnamefont {J.}~\bibnamefont
  {Aasi}} \emph {et~al.} (\bibinfo {collaboration} {LIGO Scientific}),\ }\href
  {\doibase 10.1088/0264-9381/32/7/074001} {\bibfield  {journal} {\bibinfo
  {journal} {Class. Quant. Grav.}\ }\textbf {\bibinfo {volume} {32}},\ \bibinfo
  {pages} {074001} (\bibinfo {year} {2015})},\ \Eprint
  {http://arxiv.org/abs/1411.4547} {arXiv:1411.4547 [gr-qc]} \BibitemShut
  {NoStop}%
\bibitem [{\citenamefont {Acernese}\ \emph
  {et~al.}(2015{\natexlab{b}})\citenamefont {Acernese} \emph
  {et~al.}}]{TheVirgo:2014hva}%
  \BibitemOpen
  \bibfield  {author} {\bibinfo {author} {\bibfnamefont {F.}~\bibnamefont
  {Acernese}} \emph {et~al.} (\bibinfo {collaboration} {VIRGO}),\ }\href
  {\doibase 10.1088/0264-9381/32/2/024001} {\bibfield  {journal} {\bibinfo
  {journal} {Class. Quant. Grav.}\ }\textbf {\bibinfo {volume} {32}},\ \bibinfo
  {pages} {024001} (\bibinfo {year} {2015}{\natexlab{b}})},\ \Eprint
  {http://arxiv.org/abs/1408.3978} {arXiv:1408.3978 [gr-qc]} \BibitemShut
  {NoStop}%
\bibitem [{\citenamefont {Akutsu}\ \emph {et~al.}(2019)\citenamefont {Akutsu}
  \emph {et~al.}}]{Akutsu:2018axf}%
  \BibitemOpen
  \bibfield  {author} {\bibinfo {author} {\bibfnamefont {T.}~\bibnamefont
  {Akutsu}} \emph {et~al.} (\bibinfo {collaboration} {KAGRA}),\ }\href
  {\doibase 10.1038/s41550-018-0658-y} {\bibfield  {journal} {\bibinfo
  {journal} {Nature Astron.}\ }\textbf {\bibinfo {volume} {3}},\ \bibinfo
  {pages} {35} (\bibinfo {year} {2019})},\ \Eprint
  {http://arxiv.org/abs/1811.08079} {arXiv:1811.08079 [gr-qc]} \BibitemShut
  {NoStop}%
\bibitem [{\citenamefont {Adhikari}\ \emph
  {et~al.}(2020{\natexlab{b}})\citenamefont {Adhikari} \emph
  {et~al.}}]{Adhikari:2020gft}%
  \BibitemOpen
  \bibfield  {author} {\bibinfo {author} {\bibfnamefont {R.~X.}\ \bibnamefont
  {Adhikari}} \emph {et~al.} (\bibinfo {collaboration} {LIGO}),\ }\href
  {\doibase 10.1088/1361-6382/ab9143} {\bibfield  {journal} {\bibinfo
  {journal} {Class. Quant. Grav.}\ }\textbf {\bibinfo {volume} {37}},\ \bibinfo
  {pages} {165003} (\bibinfo {year} {2020}{\natexlab{b}})},\ \Eprint
  {http://arxiv.org/abs/2001.11173} {arXiv:2001.11173 [astro-ph.IM]}
  \BibitemShut {NoStop}%
\bibitem [{\citenamefont {Abbott}\ \emph {et~al.}(2020)\citenamefont {Abbott}
  \emph {et~al.}}]{Abbott:2020gyp}%
  \BibitemOpen
  \bibfield  {author} {\bibinfo {author} {\bibfnamefont {R.}~\bibnamefont
  {Abbott}} \emph {et~al.} (\bibinfo {collaboration} {LIGO Scientific,
  Virgo}),\ }\href@noop {} {\  (\bibinfo {year} {2020})},\ \Eprint
  {http://arxiv.org/abs/2010.14533} {arXiv:2010.14533 [astro-ph.HE]}
  \BibitemShut {NoStop}%
\bibitem [{\citenamefont {Belgacem}\ \emph {et~al.}(2019)\citenamefont
  {Belgacem}, \citenamefont {Dirian}, \citenamefont {Foffa}, \citenamefont
  {Howell}, \citenamefont {Maggiore},\ and\ \citenamefont
  {Regimbau}}]{Belgacem:2019tbw}%
  \BibitemOpen
  \bibfield  {author} {\bibinfo {author} {\bibfnamefont {E.}~\bibnamefont
  {Belgacem}}, \bibinfo {author} {\bibfnamefont {Y.}~\bibnamefont {Dirian}},
  \bibinfo {author} {\bibfnamefont {S.}~\bibnamefont {Foffa}}, \bibinfo
  {author} {\bibfnamefont {E.~J.}\ \bibnamefont {Howell}}, \bibinfo {author}
  {\bibfnamefont {M.}~\bibnamefont {Maggiore}}, \ and\ \bibinfo {author}
  {\bibfnamefont {T.}~\bibnamefont {Regimbau}},\ }\href {\doibase
  10.1088/1475-7516/2019/08/015} {\bibfield  {journal} {\bibinfo  {journal}
  {JCAP}\ }\textbf {\bibinfo {volume} {08}},\ \bibinfo {pages} {015} (\bibinfo
  {year} {2019})},\ \Eprint {http://arxiv.org/abs/1907.01487} {arXiv:1907.01487
  [astro-ph.CO]} \BibitemShut {NoStop}%
\bibitem [{\citenamefont {Vangioni}\ \emph {et~al.}(2015)\citenamefont
  {Vangioni}, \citenamefont {Olive}, \citenamefont {Prestegard}, \citenamefont
  {Silk}, \citenamefont {Petitjean},\ and\ \citenamefont
  {Mandic}}]{Vangioni:2014axa}%
  \BibitemOpen
  \bibfield  {author} {\bibinfo {author} {\bibfnamefont {E.}~\bibnamefont
  {Vangioni}}, \bibinfo {author} {\bibfnamefont {K.~A.}\ \bibnamefont {Olive}},
  \bibinfo {author} {\bibfnamefont {T.}~\bibnamefont {Prestegard}}, \bibinfo
  {author} {\bibfnamefont {J.}~\bibnamefont {Silk}}, \bibinfo {author}
  {\bibfnamefont {P.}~\bibnamefont {Petitjean}}, \ and\ \bibinfo {author}
  {\bibfnamefont {V.}~\bibnamefont {Mandic}},\ }\href {\doibase
  10.1093/mnras/stu2600} {\bibfield  {journal} {\bibinfo  {journal} {Mon. Not.
  Roy. Astron. Soc.}\ }\textbf {\bibinfo {volume} {447}},\ \bibinfo {pages}
  {2575} (\bibinfo {year} {2015})},\ \Eprint {http://arxiv.org/abs/1409.2462}
  {arXiv:1409.2462 [astro-ph.GA]} \BibitemShut {NoStop}%
\bibitem [{\citenamefont {Abbott}\ \emph {et~al.}(2021)\citenamefont {Abbott}
  \emph {et~al.}}]{LIGOScientific:2021psn}%
  \BibitemOpen
  \bibfield  {author} {\bibinfo {author} {\bibfnamefont {R.}~\bibnamefont
  {Abbott}} \emph {et~al.} (\bibinfo {collaboration} {LIGO Scientific, VIRGO,
  KAGRA}),\ }\href@noop {} {\  (\bibinfo {year} {2021})},\ \Eprint
  {http://arxiv.org/abs/2111.03634} {arXiv:2111.03634 [astro-ph.HE]}
  \BibitemShut {NoStop}%
\bibitem [{\citenamefont {Kalogera}\ \emph {et~al.}(2021)\citenamefont
  {Kalogera} \emph {et~al.}}]{Kalogera:2021bya}%
  \BibitemOpen
  \bibfield  {author} {\bibinfo {author} {\bibfnamefont {V.}~\bibnamefont
  {Kalogera}} \emph {et~al.},\ }\href@noop {} {\  (\bibinfo {year} {2021})},\
  \Eprint {http://arxiv.org/abs/2111.06990} {arXiv:2111.06990 [gr-qc]}
  \BibitemShut {NoStop}%
\bibitem [{\citenamefont {Howell}\ \emph {et~al.}(2018)\citenamefont {Howell},
  \citenamefont {Ackley}, \citenamefont {Rowlinson},\ and\ \citenamefont
  {Coward}}]{Howell:2018nhu}%
  \BibitemOpen
  \bibfield  {author} {\bibinfo {author} {\bibfnamefont {E.~J.}\ \bibnamefont
  {Howell}}, \bibinfo {author} {\bibfnamefont {K.}~\bibnamefont {Ackley}},
  \bibinfo {author} {\bibfnamefont {A.}~\bibnamefont {Rowlinson}}, \ and\
  \bibinfo {author} {\bibfnamefont {D.}~\bibnamefont {Coward}},\ }\href
  {\doibase 10.1093/mnras/stz455} {\  (\bibinfo {year} {2018}),\
  10.1093/mnras/stz455},\ \Eprint {http://arxiv.org/abs/1811.09168}
  {arXiv:1811.09168 [astro-ph.HE]} \BibitemShut {NoStop}%
\bibitem [{\citenamefont {Wanderman}\ and\ \citenamefont
  {Piran}(2015)}]{Wanderman:2014eza}%
  \BibitemOpen
  \bibfield  {author} {\bibinfo {author} {\bibfnamefont {D.}~\bibnamefont
  {Wanderman}}\ and\ \bibinfo {author} {\bibfnamefont {T.}~\bibnamefont
  {Piran}},\ }\href {\doibase 10.1093/mnras/stv123} {\bibfield  {journal}
  {\bibinfo  {journal} {Mon. Not. Roy. Astron. Soc.}\ }\textbf {\bibinfo
  {volume} {448}},\ \bibinfo {pages} {3026} (\bibinfo {year} {2015})},\ \Eprint
  {http://arxiv.org/abs/1405.5878} {arXiv:1405.5878 [astro-ph.HE]} \BibitemShut
  {NoStop}%
\bibitem [{\citenamefont {Burns}\ \emph {et~al.}(2016)\citenamefont {Burns},
  \citenamefont {Connaughton}, \citenamefont {Zhang}, \citenamefont {Lien},
  \citenamefont {Briggs}, \citenamefont {Goldstein}, \citenamefont {Pelassa},\
  and\ \citenamefont {Troja}}]{Burns:2015fol}%
  \BibitemOpen
  \bibfield  {author} {\bibinfo {author} {\bibfnamefont {E.}~\bibnamefont
  {Burns}}, \bibinfo {author} {\bibfnamefont {V.}~\bibnamefont {Connaughton}},
  \bibinfo {author} {\bibfnamefont {B.-B.}\ \bibnamefont {Zhang}}, \bibinfo
  {author} {\bibfnamefont {A.}~\bibnamefont {Lien}}, \bibinfo {author}
  {\bibfnamefont {M.~S.}\ \bibnamefont {Briggs}}, \bibinfo {author}
  {\bibfnamefont {A.}~\bibnamefont {Goldstein}}, \bibinfo {author}
  {\bibfnamefont {V.}~\bibnamefont {Pelassa}}, \ and\ \bibinfo {author}
  {\bibfnamefont {E.}~\bibnamefont {Troja}},\ }\href {\doibase
  10.3847/0004-637X/818/2/110} {\bibfield  {journal} {\bibinfo  {journal}
  {Astrophys. J.}\ }\textbf {\bibinfo {volume} {818}},\ \bibinfo {pages} {110}
  (\bibinfo {year} {2016})},\ \Eprint {http://arxiv.org/abs/1512.00923}
  {arXiv:1512.00923 [astro-ph.HE]} \BibitemShut {NoStop}%
\bibitem [{\citenamefont {Gupta}\ \emph {et~al.}(2019)\citenamefont {Gupta},
  \citenamefont {Fox}, \citenamefont {Sathyaprakash},\ and\ \citenamefont
  {Schutz}}]{Gupta:2019okl}%
  \BibitemOpen
  \bibfield  {author} {\bibinfo {author} {\bibfnamefont {A.}~\bibnamefont
  {Gupta}}, \bibinfo {author} {\bibfnamefont {D.}~\bibnamefont {Fox}}, \bibinfo
  {author} {\bibfnamefont {B.~S.}\ \bibnamefont {Sathyaprakash}}, \ and\
  \bibinfo {author} {\bibfnamefont {B.~F.}\ \bibnamefont {Schutz}},\ }\href
  {\doibase 10.3847/1538-4357/ab4c92} {\bibfield  {journal} {\bibinfo
  {journal} {The Astrophysical Journal}\ }\textbf {\bibinfo {volume} {886}},\
  \bibinfo {pages} {71} (\bibinfo {year} {2019})}\BibitemShut {NoStop}%
\bibitem [{\citenamefont {{Li}}\ \emph {et~al.}(2011)\citenamefont {{Li}},
  \citenamefont {{Chornock}}, \citenamefont {{Leaman}}, \citenamefont
  {{Filippenko}}, \citenamefont {{Poznanski}}, \citenamefont {{Wang}},
  \citenamefont {{Ganeshalingam}},\ and\ \citenamefont
  {{Mannucci}}}]{2011MNRAS.412.1473L}%
  \BibitemOpen
  \bibfield  {author} {\bibinfo {author} {\bibfnamefont {W.}~\bibnamefont
  {{Li}}}, \bibinfo {author} {\bibfnamefont {R.}~\bibnamefont {{Chornock}}},
  \bibinfo {author} {\bibfnamefont {J.}~\bibnamefont {{Leaman}}}, \bibinfo
  {author} {\bibfnamefont {A.~V.}\ \bibnamefont {{Filippenko}}}, \bibinfo
  {author} {\bibfnamefont {D.}~\bibnamefont {{Poznanski}}}, \bibinfo {author}
  {\bibfnamefont {X.}~\bibnamefont {{Wang}}}, \bibinfo {author} {\bibfnamefont
  {M.}~\bibnamefont {{Ganeshalingam}}}, \ and\ \bibinfo {author} {\bibfnamefont
  {F.}~\bibnamefont {{Mannucci}}},\ }\href {\doibase
  10.1111/j.1365-2966.2011.18162.x} {\bibfield  {journal} {\bibinfo  {journal}
  {\mnras}\ }\textbf {\bibinfo {volume} {412}},\ \bibinfo {pages} {1473}
  (\bibinfo {year} {2011})},\ \Eprint {http://arxiv.org/abs/1006.4613}
  {arXiv:1006.4613 [astro-ph.SR]} \BibitemShut {NoStop}%
\bibitem [{\citenamefont {Abbott}\ \emph {et~al.}(2019)\citenamefont {Abbott}
  \emph {et~al.}}]{LIGOScientific:2018mvr}%
  \BibitemOpen
  \bibfield  {author} {\bibinfo {author} {\bibfnamefont {B.~P.}\ \bibnamefont
  {Abbott}} \emph {et~al.} (\bibinfo {collaboration} {LIGO Scientific,
  Virgo}),\ }\href {\doibase 10.1103/PhysRevX.9.031040} {\bibfield  {journal}
  {\bibinfo  {journal} {Phys. Rev. X}\ }\textbf {\bibinfo {volume} {9}},\
  \bibinfo {pages} {031040} (\bibinfo {year} {2019})},\ \Eprint
  {http://arxiv.org/abs/1811.12907} {arXiv:1811.12907 [astro-ph.HE]}
  \BibitemShut {NoStop}%
\bibitem [{\citenamefont {{Friedmann}}\ and\ \citenamefont
  {{Maoz}}(2018)}]{2018MNRAS.479.3563F}%
  \BibitemOpen
  \bibfield  {author} {\bibinfo {author} {\bibfnamefont {M.}~\bibnamefont
  {{Friedmann}}}\ and\ \bibinfo {author} {\bibfnamefont {D.}~\bibnamefont
  {{Maoz}}},\ }\href {\doibase 10.1093/mnras/sty1664} {\bibfield  {journal}
  {\bibinfo  {journal} {\mnras}\ }\textbf {\bibinfo {volume} {479}},\ \bibinfo
  {pages} {3563} (\bibinfo {year} {2018})},\ \Eprint
  {http://arxiv.org/abs/1803.04421} {arXiv:1803.04421 [astro-ph.GA]}
  \BibitemShut {NoStop}%
\bibitem [{\citenamefont {{Girardi}}\ \emph {et~al.}(2002)\citenamefont
  {{Girardi}}, \citenamefont {{Manzato}}, \citenamefont {{Mezzetti}},
  \citenamefont {{Giuricin}},\ and\ \citenamefont
  {{Limboz}}}]{Girardi:2002mmg}%
  \BibitemOpen
  \bibfield  {author} {\bibinfo {author} {\bibfnamefont {M.}~\bibnamefont
  {{Girardi}}}, \bibinfo {author} {\bibfnamefont {P.}~\bibnamefont
  {{Manzato}}}, \bibinfo {author} {\bibfnamefont {M.}~\bibnamefont
  {{Mezzetti}}}, \bibinfo {author} {\bibfnamefont {G.}~\bibnamefont
  {{Giuricin}}}, \ and\ \bibinfo {author} {\bibfnamefont {F.}~\bibnamefont
  {{Limboz}}},\ }\href {\doibase 10.1086/339360} {\bibfield  {journal}
  {\bibinfo  {journal} {\apj}\ }\textbf {\bibinfo {volume} {569}},\ \bibinfo
  {pages} {720} (\bibinfo {year} {2002})},\ \Eprint
  {http://arxiv.org/abs/astro-ph/0112534} {astro-ph/0112534} \BibitemShut
  {NoStop}%
\bibitem [{\citenamefont {Aghanim}\ \emph {et~al.}(2020)\citenamefont {Aghanim}
  \emph {et~al.}}]{Aghanim:2018eyx}%
  \BibitemOpen
  \bibfield  {author} {\bibinfo {author} {\bibfnamefont {N.}~\bibnamefont
  {Aghanim}} \emph {et~al.} (\bibinfo {collaboration} {Planck}),\ }\href
  {\doibase 10.1051/0004-6361/201833910} {\bibfield  {journal} {\bibinfo
  {journal} {Astron. Astrophys.}\ }\textbf {\bibinfo {volume} {641}},\ \bibinfo
  {pages} {A6} (\bibinfo {year} {2020})},\ \Eprint
  {http://arxiv.org/abs/1807.06209} {arXiv:1807.06209 [astro-ph.CO]}
  \BibitemShut {NoStop}%
\bibitem [{\citenamefont {Borhanian}(2020)}]{Borhanian:2020ypi}%
  \BibitemOpen
  \bibfield  {author} {\bibinfo {author} {\bibfnamefont {S.}~\bibnamefont
  {Borhanian}},\ }\href@noop {} {\  (\bibinfo {year} {2020})},\ \Eprint
  {http://arxiv.org/abs/2010.15202} {arXiv:2010.15202 [gr-qc]} \BibitemShut
  {NoStop}%
\bibitem [{\citenamefont {Cutler}\ and\ \citenamefont
  {Flanagan}(1994)}]{Cutler:1994ys}%
  \BibitemOpen
  \bibfield  {author} {\bibinfo {author} {\bibfnamefont {C.}~\bibnamefont
  {Cutler}}\ and\ \bibinfo {author} {\bibfnamefont {E.~E.}\ \bibnamefont
  {Flanagan}},\ }\href {\doibase 10.1103/PhysRevD.49.2658} {\bibfield
  {journal} {\bibinfo  {journal} {Phys. Rev. D}\ }\textbf {\bibinfo {volume}
  {49}},\ \bibinfo {pages} {2658} (\bibinfo {year} {1994})},\ \Eprint
  {http://arxiv.org/abs/gr-qc/9402014} {arXiv:gr-qc/9402014} \BibitemShut
  {NoStop}%
\bibitem [{\citenamefont {Poisson}\ and\ \citenamefont
  {Will}(1995)}]{Poisson:1995ef}%
  \BibitemOpen
  \bibfield  {author} {\bibinfo {author} {\bibfnamefont {E.}~\bibnamefont
  {Poisson}}\ and\ \bibinfo {author} {\bibfnamefont {C.~M.}\ \bibnamefont
  {Will}},\ }\href {\doibase 10.1103/PhysRevD.52.848} {\bibfield  {journal}
  {\bibinfo  {journal} {Phys. Rev. D}\ }\textbf {\bibinfo {volume} {52}},\
  \bibinfo {pages} {848} (\bibinfo {year} {1995})},\ \Eprint
  {http://arxiv.org/abs/gr-qc/9502040} {arXiv:gr-qc/9502040} \BibitemShut
  {NoStop}%
\bibitem [{\citenamefont {{Markovi{\'c}}}(1993)}]{1993PhRvD..48.4738M}%
  \BibitemOpen
  \bibfield  {author} {\bibinfo {author} {\bibfnamefont {D.}~\bibnamefont
  {{Markovi{\'c}}}},\ }\href {\doibase 10.1103/PhysRevD.48.4738} {\bibfield
  {journal} {\bibinfo  {journal} {\prd}\ }\textbf {\bibinfo {volume} {48}},\
  \bibinfo {pages} {4738} (\bibinfo {year} {1993})}\BibitemShut {NoStop}%
\bibitem [{\citenamefont {{Amanullah}}\ \emph {et~al.}(2010)\citenamefont
  {{Amanullah}}, \citenamefont {{Lidman}}, \citenamefont {{Rubin}},
  \citenamefont {{Aldering}}, \citenamefont {{Astier}}, \citenamefont
  {{Barbary}}, \citenamefont {{Burns}}, \citenamefont {{Conley}}, \citenamefont
  {{Dawson}}, \citenamefont {{Deustua}}, \citenamefont {{Doi}}, \citenamefont
  {{Fabbro}}, \citenamefont {{Faccioli}}, \citenamefont {{Fakhouri}},
  \citenamefont {{Folatelli}}, \citenamefont {{Fruchter}}, \citenamefont
  {{Furusawa}}, \citenamefont {{Garavini}}, \citenamefont {{Goldhaber}},
  \citenamefont {{Goobar}}, \citenamefont {{Groom}}, \citenamefont {{Hook}},
  \citenamefont {{Howell}}, \citenamefont {{Kashikawa}}, \citenamefont {{Kim}},
  \citenamefont {{Knop}}, \citenamefont {{Kowalski}}, \citenamefont {{Linder}},
  \citenamefont {{Meyers}}, \citenamefont {{Morokuma}}, \citenamefont
  {{Nobili}}, \citenamefont {{Nordin}}, \citenamefont {{Nugent}}, \citenamefont
  {{{\"O}stman}}, \citenamefont {{Pain}}, \citenamefont {{Panagia}},
  \citenamefont {{Perlmutter}}, \citenamefont {{Raux}}, \citenamefont
  {{Ruiz-Lapuente}}, \citenamefont {{Spadafora}}, \citenamefont {{Strovink}},
  \citenamefont {{Suzuki}}, \citenamefont {{Wang}}, \citenamefont
  {{Wood-Vasey}}, \citenamefont {{Yasuda}},\ and\ \citenamefont {{Supernova
  Cosmology Project}}}]{2010ApJ...716..712A}%
  \BibitemOpen
  \bibfield  {author} {\bibinfo {author} {\bibfnamefont {R.}~\bibnamefont
  {{Amanullah}}}, \bibinfo {author} {\bibfnamefont {C.}~\bibnamefont
  {{Lidman}}}, \bibinfo {author} {\bibfnamefont {D.}~\bibnamefont {{Rubin}}},
  \bibinfo {author} {\bibfnamefont {G.}~\bibnamefont {{Aldering}}}, \bibinfo
  {author} {\bibfnamefont {P.}~\bibnamefont {{Astier}}}, \bibinfo {author}
  {\bibfnamefont {K.}~\bibnamefont {{Barbary}}}, \bibinfo {author}
  {\bibfnamefont {M.~S.}\ \bibnamefont {{Burns}}}, \bibinfo {author}
  {\bibfnamefont {A.}~\bibnamefont {{Conley}}}, \bibinfo {author}
  {\bibfnamefont {K.~S.}\ \bibnamefont {{Dawson}}}, \bibinfo {author}
  {\bibfnamefont {S.~E.}\ \bibnamefont {{Deustua}}}, \bibinfo {author}
  {\bibfnamefont {M.}~\bibnamefont {{Doi}}}, \bibinfo {author} {\bibfnamefont
  {S.}~\bibnamefont {{Fabbro}}}, \bibinfo {author} {\bibfnamefont
  {L.}~\bibnamefont {{Faccioli}}}, \bibinfo {author} {\bibfnamefont {H.~K.}\
  \bibnamefont {{Fakhouri}}}, \bibinfo {author} {\bibfnamefont
  {G.}~\bibnamefont {{Folatelli}}}, \bibinfo {author} {\bibfnamefont {A.~S.}\
  \bibnamefont {{Fruchter}}}, \bibinfo {author} {\bibfnamefont
  {H.}~\bibnamefont {{Furusawa}}}, \bibinfo {author} {\bibfnamefont
  {G.}~\bibnamefont {{Garavini}}}, \bibinfo {author} {\bibfnamefont
  {G.}~\bibnamefont {{Goldhaber}}}, \bibinfo {author} {\bibfnamefont
  {A.}~\bibnamefont {{Goobar}}}, \bibinfo {author} {\bibfnamefont {D.~E.}\
  \bibnamefont {{Groom}}}, \bibinfo {author} {\bibfnamefont {I.}~\bibnamefont
  {{Hook}}}, \bibinfo {author} {\bibfnamefont {D.~A.}\ \bibnamefont
  {{Howell}}}, \bibinfo {author} {\bibfnamefont {N.}~\bibnamefont
  {{Kashikawa}}}, \bibinfo {author} {\bibfnamefont {A.~G.}\ \bibnamefont
  {{Kim}}}, \bibinfo {author} {\bibfnamefont {R.~A.}\ \bibnamefont {{Knop}}},
  \bibinfo {author} {\bibfnamefont {M.}~\bibnamefont {{Kowalski}}}, \bibinfo
  {author} {\bibfnamefont {E.}~\bibnamefont {{Linder}}}, \bibinfo {author}
  {\bibfnamefont {J.}~\bibnamefont {{Meyers}}}, \bibinfo {author}
  {\bibfnamefont {T.}~\bibnamefont {{Morokuma}}}, \bibinfo {author}
  {\bibfnamefont {S.}~\bibnamefont {{Nobili}}}, \bibinfo {author}
  {\bibfnamefont {J.}~\bibnamefont {{Nordin}}}, \bibinfo {author}
  {\bibfnamefont {P.~E.}\ \bibnamefont {{Nugent}}}, \bibinfo {author}
  {\bibfnamefont {L.}~\bibnamefont {{{\"O}stman}}}, \bibinfo {author}
  {\bibfnamefont {R.}~\bibnamefont {{Pain}}}, \bibinfo {author} {\bibfnamefont
  {N.}~\bibnamefont {{Panagia}}}, \bibinfo {author} {\bibfnamefont
  {S.}~\bibnamefont {{Perlmutter}}}, \bibinfo {author} {\bibfnamefont
  {J.}~\bibnamefont {{Raux}}}, \bibinfo {author} {\bibfnamefont
  {P.}~\bibnamefont {{Ruiz-Lapuente}}}, \bibinfo {author} {\bibfnamefont
  {A.~L.}\ \bibnamefont {{Spadafora}}}, \bibinfo {author} {\bibfnamefont
  {M.}~\bibnamefont {{Strovink}}}, \bibinfo {author} {\bibfnamefont
  {N.}~\bibnamefont {{Suzuki}}}, \bibinfo {author} {\bibfnamefont
  {L.}~\bibnamefont {{Wang}}}, \bibinfo {author} {\bibfnamefont {W.~M.}\
  \bibnamefont {{Wood-Vasey}}}, \bibinfo {author} {\bibfnamefont
  {N.}~\bibnamefont {{Yasuda}}}, \ and\ \bibinfo {author} {\bibfnamefont
  {T.}~\bibnamefont {{Supernova Cosmology Project}}},\ }\href {\doibase
  10.1088/0004-637X/716/1/712} {\bibfield  {journal} {\bibinfo  {journal}
  {\apj}\ }\textbf {\bibinfo {volume} {716}},\ \bibinfo {pages} {712} (\bibinfo
  {year} {2010})},\ \Eprint {http://arxiv.org/abs/1004.1711} {arXiv:1004.1711
  [astro-ph.CO]} \BibitemShut {NoStop}%
\bibitem [{LSS()}]{LSST:SNeIa}%
  \BibitemOpen
  \href@noop {} {}\bibinfo {note} {{11.5 Constraining the Dark Energy Equation
  of State,
  \url{https://www.lsst.org/sites/default/files/docs/sciencebook/SB_11.pdf}}}\BibitemShut
  {NoStop}%
\bibitem [{\citenamefont {Riess}\ \emph {et~al.}(2021)\citenamefont {Riess}
  \emph {et~al.}}]{Riess:2021jrx}%
  \BibitemOpen
  \bibfield  {author} {\bibinfo {author} {\bibfnamefont {A.~G.}\ \bibnamefont
  {Riess}} \emph {et~al.},\ }\href@noop {} {\  (\bibinfo {year} {2021})},\
  \Eprint {http://arxiv.org/abs/2112.04510} {arXiv:2112.04510 [astro-ph.CO]}
  \BibitemShut {NoStop}%
\bibitem [{\citenamefont {Kashyap}\ \emph {et~al.}(2019)\citenamefont
  {Kashyap}, \citenamefont {Raman},\ and\ \citenamefont
  {Ajith}}]{Kashyap:2019ypm}%
  \BibitemOpen
  \bibfield  {author} {\bibinfo {author} {\bibfnamefont {R.}~\bibnamefont
  {Kashyap}}, \bibinfo {author} {\bibfnamefont {G.}~\bibnamefont {Raman}}, \
  and\ \bibinfo {author} {\bibfnamefont {P.}~\bibnamefont {Ajith}},\ }\href
  {\doibase 10.3847/2041-8213/ab543f} {\bibfield  {journal} {\bibinfo
  {journal} {Astrophys. J. Lett.}\ }\textbf {\bibinfo {volume} {886}},\
  \bibinfo {pages} {L19} (\bibinfo {year} {2019})},\ \Eprint
  {http://arxiv.org/abs/1908.02168} {arXiv:1908.02168 [astro-ph.SR]}
  \BibitemShut {NoStop}%
\bibitem [{\citenamefont {Coughlin}\ \emph {et~al.}(2020)\citenamefont
  {Coughlin}, \citenamefont {Dietrich}, \citenamefont {Heinzel}, \citenamefont
  {Khetan}, \citenamefont {Antier}, \citenamefont {Bulla}, \citenamefont
  {Christensen}, \citenamefont {Coulter},\ and\ \citenamefont
  {Foley}}]{Coughlin:2019vtv}%
  \BibitemOpen
  \bibfield  {author} {\bibinfo {author} {\bibfnamefont {M.~W.}\ \bibnamefont
  {Coughlin}}, \bibinfo {author} {\bibfnamefont {T.}~\bibnamefont {Dietrich}},
  \bibinfo {author} {\bibfnamefont {J.}~\bibnamefont {Heinzel}}, \bibinfo
  {author} {\bibfnamefont {N.}~\bibnamefont {Khetan}}, \bibinfo {author}
  {\bibfnamefont {S.}~\bibnamefont {Antier}}, \bibinfo {author} {\bibfnamefont
  {M.}~\bibnamefont {Bulla}}, \bibinfo {author} {\bibfnamefont
  {N.}~\bibnamefont {Christensen}}, \bibinfo {author} {\bibfnamefont {D.~A.}\
  \bibnamefont {Coulter}}, \ and\ \bibinfo {author} {\bibfnamefont {R.~J.}\
  \bibnamefont {Foley}},\ }\href {\doibase 10.1103/PhysRevResearch.2.022006}
  {\bibfield  {journal} {\bibinfo  {journal} {Phys. Rev. Res.}\ }\textbf
  {\bibinfo {volume} {2}},\ \bibinfo {pages} {022006} (\bibinfo {year}
  {2020})},\ \Eprint {http://arxiv.org/abs/1908.00889} {arXiv:1908.00889
  [astro-ph.HE]} \BibitemShut {NoStop}%
\bibitem [{\citenamefont {Abbott}\ \emph
  {et~al.}(2017{\natexlab{c}})\citenamefont {Abbott} \emph
  {et~al.}}]{Abbott:2017xzu}%
  \BibitemOpen
  \bibfield  {author} {\bibinfo {author} {\bibfnamefont {B.~P.}\ \bibnamefont
  {Abbott}} \emph {et~al.} (\bibinfo {collaboration} {LIGO Scientific, Virgo,
  1M2H, Dark Energy Camera GW-E, DES, DLT40, Las Cumbres Observatory, VINROUGE,
  MASTER}),\ }\href {\doibase 10.1038/nature24471} {\bibfield  {journal}
  {\bibinfo  {journal} {Nature}\ }\textbf {\bibinfo {volume} {551}},\ \bibinfo
  {pages} {85} (\bibinfo {year} {2017}{\natexlab{c}})},\ \Eprint
  {http://arxiv.org/abs/1710.05835} {arXiv:1710.05835 [astro-ph.CO]}
  \BibitemShut {NoStop}%
\bibitem [{\citenamefont {Kasen}\ \emph {et~al.}(2017)\citenamefont {Kasen},
  \citenamefont {Metzger}, \citenamefont {Barnes}, \citenamefont {Quataert},\
  and\ \citenamefont {Ramirez-Ruiz}}]{Kasen:2017sxr}%
  \BibitemOpen
  \bibfield  {author} {\bibinfo {author} {\bibfnamefont {D.}~\bibnamefont
  {Kasen}}, \bibinfo {author} {\bibfnamefont {B.}~\bibnamefont {Metzger}},
  \bibinfo {author} {\bibfnamefont {J.}~\bibnamefont {Barnes}}, \bibinfo
  {author} {\bibfnamefont {E.}~\bibnamefont {Quataert}}, \ and\ \bibinfo
  {author} {\bibfnamefont {E.}~\bibnamefont {Ramirez-Ruiz}},\ }\href {\doibase
  10.1038/nature24453} {\bibfield  {journal} {\bibinfo  {journal} {Nature}\
  }\textbf {\bibinfo {volume} {551}},\ \bibinfo {pages} {80} (\bibinfo {year}
  {2017})},\ \Eprint {http://arxiv.org/abs/1710.05463} {arXiv:1710.05463
  [astro-ph.HE]} \BibitemShut {NoStop}%
\bibitem [{\citenamefont {Bulla}(2019)}]{Bulla:2019muo}%
  \BibitemOpen
  \bibfield  {author} {\bibinfo {author} {\bibfnamefont {M.}~\bibnamefont
  {Bulla}},\ }\href {\doibase 10.1093/mnras/stz2495} {\bibfield  {journal}
  {\bibinfo  {journal} {Mon. Not. Roy. Astron. Soc.}\ }\textbf {\bibinfo
  {volume} {489}},\ \bibinfo {pages} {5037} (\bibinfo {year} {2019})},\ \Eprint
  {http://arxiv.org/abs/1906.04205} {arXiv:1906.04205 [astro-ph.HE]}
  \BibitemShut {NoStop}%
\end{thebibliography}%
\end{document}